\documentclass[aps,prd,10pt,nofootinbib,OliveGeqsecnum,showpacs,showkeys,superscriptaddress,preprintnumbers,altaffilletter,floatfix,twocolumn]{revtex4-2}

\usepackage{graphicx,caption,subcaption,bm,color,hyperref}
\usepackage{amsmath,amssymb}
\usepackage{mathtools} 
\usepackage{dsfont} 
\usepackage{multirow} 
\usepackage{float} 
\usepackage{mathrsfs} 
\usepackage{tensor} 
\usepackage{booktabs}
\usepackage{caption}
\usepackage{ragged2e}
\usepackage{caption}
\usepackage{tabularx} 
\usepackage[british]{babel}
\usepackage[table]{xcolor}
\usepackage{array}

\hypersetup{colorlinks, linkcolor={blue}, citecolor={red}, urlcolor={teal}}

\definecolor{amaranth}{rgb}{0.9, 0.17, 0.31}
\definecolor{palatinateblue}{rgb}{0.15, 0.23, 0.89}
\definecolor{brightpink}{rgb}{1.0, 0.0, 0.5}
\definecolor{brightgreen}{rgb}{0.14, 0.84, 0.72}
\definecolor{mediumgreen}{rgb}{0.22, 0.67, 0.59}
\definecolor{darkgreen}{rgb}{0.25, 0.5, 0.46}
\definecolor{verydarkgreen}{rgb}{0.22, 0.33, 0.31}

\hypersetup{ linktoc=all,
    colorlinks, linkcolor={mediumgreen},
    citecolor={mediumgreen}, urlcolor={darkgreen}
}
\DeclareUnicodeCharacter{03A9}{$\Omega$}
\DeclareUnicodeCharacter{039B}{$\Lambda$}

\graphicspath{{images/}}

\newcommand{\be}{\begin{equation}}
\newcommand{\ee}{\end{equation}}
\newcommand{\ba}{\begin{eqnarray}}
\newcommand{\ea}{\end{eqnarray}}

\begin{document}

\title{Sign-Switching Dark Energy: Smooth Transitions with Recent \textit{DESI DR2} Observations}

\author{Beñat Ibarra-Uriondo}
\email{benat.ibarra@ehu.eus}
\affiliation{Department of Physics, University of the Basque Country EHU, P.O. Box 644, 48080 Bilbao, Spain}
\affiliation{EHU Quantum Center, University of the Basque Country EHU, P.O. Box 644, 48080 Bilbao, Spain}

\author{Mariam Bouhmadi-López}
\email{mariam.bouhmadi@ehu.eus}
\affiliation{IKERBASQUE, Basque Foundation for Science, 48011, Bilbao, Spain}
\affiliation{Department of Physics, University of the Basque Country EHU, P.O. Box 644, 48080 Bilbao, Spain}
\affiliation{EHU Quantum Center, University of the Basque Country EHU, P.O. Box 644, 48080 Bilbao, Spain}

\begin{abstract} 
Sign-switching dark energy provides a novel mechanism for modifying the late-time expansion history of the Universe without invoking additional fields or finely tuned initial conditions. In this work, we investigate a class of background--level cosmological models in which the dark energy contribution changes sign at a transition redshift $z_\dagger$, producing a sharp deviation from standard $\Lambda$CDM dynamics. We confront these models with a comprehensive set of cosmological observations, including compressed \textit{Planck 18} cosmic microwave background (CMB) measurements, \textit{DESI DR2} Baryonic Acoustic Oscillation (BAO) data and the \textit{Pantheon+} $\&$ \textit{SH0ES} Type Ia supernova sample (SN). Using a full Markov Chain Monte Carlo (MCMC) analysis, we find that the sign-switching scenario significantly alleviates the Hubble tension while obtaining better results when statistically comparing with $\Lambda$CDM, as quantified by the Akaike and Bayesian information Criteria. Although the model is explored only at the background level, the improvement in the inferred Hubble constant demonstrates that sign-switching dark energy offers a promising and physically economical pathway toward resolving late-universe discrepancies. 
\end{abstract}

\maketitle


\section{Introduction}

The late-time acceleration of the Universe, first discovered in 1998 through observations of Type Ia supernovae (SN)
\cite{SupernovaSearchTeam:1998fmf,SupernovaCosmologyProject:1998vns}, has since been firmly established by a variety 
of independent probes, including baryon acoustic oscillations (BAO) \cite{Eisenstein2005,Cole2005,Percival2010,Tegmark2004,Reid2010}, cosmic microwave background (CMB) anisotropies \cite{Bennett2003,Komatsu2011,Planck2013,Planck:2018vyg}, 
and weak gravitational lensing \cite{DES:2026zjp}. Within the Standard Model of Cosmology, this accelerated 
expansion is attributed to a positive cosmological constant $\Lambda$, giving rise to the $\Lambda$CDM model. 
While $\Lambda$CDM provides an excellent description of both the background expansion and the growth of cosmic 
structure, it faces several theoretical and observational challenges.

From a theoretical standpoint, the introduction of a constant vacuum energy density, 
$\rho_\Lambda \approx 10^{-47}~\text{GeV}^4$, poses the well-known cosmological constant problem 
\cite{RevModPhys.61.1,Sahni:1999gb}. Additionally, the coincidence problem 
\cite{Carroll:2000fy,Padmanabhan:2002ji} highlights the lack of a natural explanation for why the present-day 
energy densities of matter and dark energy (DE) are of the same order, 
$\Omega_{\rm m0}\sim\Omega_{\rm d0}$, without significant fine-tuning of initial conditions. 
To address these open issues, a wide range of alternatives to the standard $\Lambda$CDM paradigm have been proposed in the literature. On the one hand, dynamical DE models replace the cosmological constant with additional degrees of freedom that evolve in time. Prominent examples include canonical scalar field models such as quintessence \cite{PhysRevD.37.3406}, non-canonical constructions such as $k$-essence \cite{PhysRevD.63.103510}, and axion-like DE fields, which are well motivated by high-energy physics and string-inspired scenarios \cite{Kamionkowski:2014zda,Emami:2016mrt,Chiang:2025qxg}. Alternative phenomenological approaches based on effective perfect-fluid descriptions have also been explored as a means of capturing departures from a pure cosmological constant at late times \cite{Kamenshchik:2001cp}. In addition to scalar field models \cite{Rezazadeh:2022lsf}, three-form fields have been shown to provide viable dynamical sources of cosmic acceleration \cite{Koivisto:2009fb,Koivisto:2009ew,Morais:2016bev,Bouhmadi-Lopez:2016dzw}, arising naturally in higher-dimensional theories and supergravity, and capable of driving late-time acceleration for suitable choices of the self-interaction potential \cite{Bouhmadi-Lopez:2025lzm}. Interacting dark energy (IDE) models featuring a sign-changing coupling constant have likewise been shown to alleviate cosmological tensions \cite{Sabogal:2025qhz,Li:2025owk,Li:2026xaz}. On the other hand, the observed cosmic acceleration may signal a breakdown of General Relativity on cosmological scales, motivating a broad class of modified gravity theories. These include extensions with additional spin-2 degrees of freedom such as bigravity \cite{Kobayashi:2019hrl}, as well as scalar–tensor frameworks featuring non-trivial derivative interactions, notably kinetic gravity braiding (KGB) models \cite{Deffayet:2010qz,Pujolas:2011he,BorislavovVasilev:2022gpp,BorislavovVasilev:2024loq}. Further well-studied examples comprise curvature-based modifications such as $f(\mathcal{R})$ gravity \cite{Sotiriou:2008rp,Capozziello:2011et,Nojiri:2010wj,Nojiri:2017ncd}, torsion-based extensions within the teleparallel framework, commonly referred to as $f(\mathcal{T})$ gravity \cite{Bengochea:2008gz,Ferraro:2006jd,Cai:2015emx,Bouhmadi-Lopez:2026dte}, and more recently, theories formulated in terms of spacetime non-metricity, including $f(\mathcal{Q})$ gravity \cite{BeltranJimenez:2018vdo,BeltranJimenez:2019tme,Ayuso:2020dcu,Boiza:2025xpn,Ayuso:2025vkc}. More general constructions can be achieved by introducing appropriate boundary terms that connect torsional and non-metric formulations of gravity, giving rise to unified extensions such as $F(Q,\mathcal{B})$ \cite{Capozziello:2023vne,De:2023xua}, $f(T,\mathcal{B})$ \cite{Bahamonde:2015zma,Bahamonde:2019shr} and $f(\mathcal{R},T)$ \cite{Lamaaoune:2025zhd} theories .

Despite its remarkable success, the $\Lambda$CDM model continues to face increasing tension with high-precision observations that probe the cosmic expansion history.
The most prominent of these is the well-known tension in the Hubble constant $H_0$ \cite{DiValentino:2021izs,Poulin:2018,Kamionkowski:2022pkx,DiValentino:2026uua,DiValentino:2026uua}, arising from the mismatch between early-Universe inferences based on CMB measurements by $\textit{Planck}$ satellite \cite{Planck:2018vyg} and late-time determinations from Type Ia supernovae calibrated via local distance-ladder techniques, most notably by the SH0ES collaboration using the Pantheon+ sample \cite{Riess:2020fzl,Brout:2022vxf}. This discrepancy, now exceeding the $5\sigma$ level, has proven robust against known systematic effects and has prompted growing scrutiny of the assumptions underlying the standard cosmological framework. At the same time, BAO observations provide an independent and highly stable probe of the background expansion, anchoring distance measurements over a broad redshift range. Recent BAO results from the dark energy Spectroscopic Instrument, particularly those from \textit{DESI DR2} \cite{DESI:2025zgx}, have significantly improved the precision of these measurements and have begun to reveal subtle preferences in combined analyses for departures from a strictly constant DE equation of state (EoS). In particular, some data combinations mildly favor an evolution of the effective EoS that approaches or crosses the cosmological constant line, $w=-1$, at late times. Although such indications remain statistically modest and compatible with $\Lambda$CDM within current uncertainties, they underscore the importance of systematically testing extensions of the standard model using background observables alone, which now reach a level of precision capable of exposing small but potentially meaningful deviations from the cosmological constant paradigm.

In this paper, we focus on an intriguing class of dynamical late-time modifications \cite{Vagnozzi:2023nrq}, which is represented by \emph{sign-switching dark energy} models, inspired by 
the graduated dark energy (gDE) framework \cite{Akarsu:2019hmw,Acquaviva:2021jov}, an earlier DE model which assumed a negative cosmological constant on top of a positive DE component can be found in \cite{Visinelli:2019qqu}. 
In these models, the effective cosmological constant undergoes a transition from negative to positive values at 
late times ($z\lesssim2$), thus interpolating between an anti–de Sitter (AdS)-like state at earlier epochs and a 
de Sitter (dS)-like state at late times. The simplest realisation, the abrupt $\Lambda_{\rm s}$CDM model 
\cite{Akarsu:2021fol,Akarsu:2022typ,Akarsu:2023mfb,Toda:2024ncp,Paraskevas:2024ytz,Akarsu:2024qsi,Akarsu:2024eoo,Akarsu:2024nas,Souza:2024qwd,Akarsu:2025gwi,Akarsu:2025ijk,Escamilla:2025imi,Akarsu:2025dmj,Akarsu:2025nns,Ghafari:2025eql,DiGennaro:2022ykp,Yadav:2025vpx,Tamayo:2025xci}, incorporates this idea through an abrupt sign flip of $\Lambda$, and has been 
shown to alleviate the $H_0$ and $S_8$ tensions \cite{Akarsu:2023mfb,CosmoVerse:2025txj,Souza:2024qwd,Escamilla:2025imi,Gomez-Valent:2024tdb,Gomez-Valent:2024ejh}. 
The presence of a negative cosmological constant is theoretically well motivated, appearing naturally in string theory, 
supergravity, and the AdS/CFT correspondence \cite{Maldacena:1997re,Wang:2026kbg,Chakraborty:2025yuo,Wang:2025dtk}. 
Moreover, constructing quantum field theory or vacuum solutions in a dS background ($\Lambda>0$) is notoriously 
difficult \cite{Obied:2018sgi,Maldacena:2000mw,Kachru:2003aw}, further encouraging models in which the Universe 
experienced an AdS-like phase at intermediate redshift. Although the sign-switching mechanism leaves the early Universe largely unaffected; since DE is subdominant, the 
transition around the matter–dark energy equality yields phenomenologically significant effects. 
Current observations typically favour transition redshifts $z_\dagger\sim 1.7$ to solve $H_0$ tension 
\cite{Akarsu:2019hmw,Akarsu:2021fol,Escamilla:2025imi,Pedrotti:2025ccw,Camlibel:2025afv}. This picture finds intriguing support in recent model-agnostic reconstructions of cosmic expansion, which have reported hints of a sign-changing scalar field energy density and a transient acceleration phase around\footnote{A very recent discussion of the null energy condition and the crossing of the cosmological constant line in sign-switching models can be found in \cite{Akarsu:2026anp}; see also Fig.~1 of \cite{Bouhmadi-Lopez:2019zvz}. As is well known, phantom dark energy — that is, a DE density that increases in an expanding Universe — does not necessarily imply an EoS parameter smaller than minus one unless the energy density is strictly positive. Given that most of the literature assumes a positive DE density, this may lead to the misconception that $1+w$ is always negative in phantom DE models. However, this is not the case in sign-switching scenarios, where phantom behaviour arises and a more careful assessment of the energy conditions is therefore required. Further discussion of the energy conditions in \textit{standard} DE models can be found in \cite{Bouhmadi-Lopez:2019zvz}.} $z\sim 2$ \cite{Akarsu:2026anp}. Furthermore, the crossing of the cosmological constant line — a feature that naturally emerges in smooth sign-switching models — has also been shown to arise in specific higher-dimensional constructions, such as dilatonic brane-world scenarios, where a genuine and smooth crossing of the cosmological constant line occurs. In these cases, the dark energy EoS parameter crosses the cosmological constant line while the sign of the energy density remains unchanged \cite{Bouhmadi-Lopez:2008ukq}.

As mentioned, the simplest realisation of sign-switching DE is characterised by a signum function (sgn) where the DE density changes sign at redshift $z_{\dagger}$. This phenomenological and simple model allows for the study of such behaviour without assuming the difficulties that come with dynamical DE models:

\begin{equation}
\Lambda\quad \rightarrow \quad \Lambda_\mathrm{s}(z) = \Lambda \text{ sgn}(z_{\dagger} - z).
\end{equation}
Subsequent developments have explored natural realisations of the sign flip.  
A notable example is the $\Lambda_{\rm s}$CDM$^{+}$ model 
\cite{Anchordoqui:2023woo,Anchordoqui:2024gfa,Anchordoqui:2024dqc,Soriano:2025gxd}, where the negative-to-positive 
transition emerges dynamically via Casimir forces acting in the bulk of a higher-dimensional space-time. 
This demonstrates how revisiting well-known theories within broader frameworks can uncover previously overlooked 
solution spaces capable of realising such cosmological transitions.

While the abrupt $\Lambda_{\rm s}$CDM implementation offers a useful phenomenological prototype, its instantaneous 
transition introduces a type-II (sudden) singularity at $z=z_\dagger$ \cite{Akarsu:2025gwi}, whose influence on structure 
formation is mild but conceptually undesirable. A more realistic formulation requires a smooth, continuous transition, 
which replaces the sudden singularity with a milder $w$-singularity \cite{Paraskevas:2024ytz,Bouhmadi-Lopez:2025ggl}. 
Motivated by the smooth transition featured in models such as $\Lambda_{\rm s}$VCDM 
\cite{Akarsu:2024qsi,Akarsu:2024eoo}, Ref.~\cite{Bouhmadi-Lopez:2025ggl,Bouhmadi-Lopez:2025spo} introduced three \textit{smooth sign-switching DE} 
extensions: a ladder-like model, a smooth-step function model, and an interpolating error-function-based model. 
These scenarios provide more physically plausible realisations of the AdS-to-dS transition and avoid the mathematical 
pathologies present in the abrupt counterpart. These smoother transitions lead to a dynamical equation-of-state parameter capable of exhibiting a late-time crossing of the cosmological constant line, in line with the mild indications emerging from recent DESI data.

In this work, we present an observational analysis of a class of sign-switching dark energy models, including both the abrupt $\Lambda_{\rm s}$CDM scenario and its smooth extensions. The latter, comprising the ladder-like, smooth-step, and error-function parametrisations, were introduced in our previous works~\cite{Bouhmadi-Lopez:2025ggl,Bouhmadi-Lopez:2025spo}, where their theoretical consistency and phenomenological properties were established. Here, we build upon that framework by confronting these models with current cosmological data, with particular emphasis on the recent \textit{DESI DR2} BAO measurements. Our analysis is restricted to background-level observables, allowing for a clean assessment of their impact on the expansion history without introducing assumptions about perturbations. The effective equation of state exhibits a divergence when the energy density crosses zero, a characteristic feature of these scenarios that we consistently account for in the analysis.

We perform Markov Chain Monte Carlo (MCMC) analyses using a combination of early- and late-time distance probes. We place observational constraints on these scenarios by performing Markov Chain Monte Carlo (MCMC) analyses using a combination of early- and late-time distance probes. In particular, we employ the \textit{Planck 2018} compressed likelihood, Type~Ia supernovae from the \textit{Pantheon+} compilation calibrated with \textit{SH0ES} local distance-ladder measurements, BAO data from the recent \textit{DESI DR2} release, and a \textit{Big Bang Nucleosynthesis} (BBN) prior. This combination of data sets provides a powerful and complementary probe of the cosmic expansion history over a wide redshift range. From the resulting posterior distributions, we derive constraints on both cosmological and DE parameters, together with reconstructions of key background quantities, including the Hubble expansion rate and selected cosmographic parameters. In addition, we assess the statistical performance of the models relative to $\Lambda$CDM using standard information criteria, thereby enabling a quantitative comparison that balances goodness of fit against model complexity.

The paper is organised as follows. In Sec.\ref{models}, we introduce the sign-switching DE models considered in this work, briefly describing the smooth transition scenarios. Sec.\ref{sec3} is devoted to the observational analysis: Sec.\ref{sec3a} presents the datasets and methodology employed in the MCMC analysis;  and Sec.\ref{sec3b} outlines the information criteria used for statistical inference and model comparison. Sec.~\ref{sec3c} discusses the resulting constraints, background reconstructions, and statistical outcomes. Finally, Sec.\ref{sec4} summarises our findings and presents our conclusions.

\section{Smooth transitions\label{models}}

Throughout this work, we assume a homogeneous and isotropic Universe described by the Friedmann-Lemaître-Robertson-Walker (FLRW) metric. Spatial flatness is imposed, and the cosmic energy budget is modelled as a collection of non-interacting fluids, including radiation, pressureless matter (comprising cold dark matter and baryons) and DE. Within this framework, we introduce a class of DE models featuring transitions between distinct regimes. We consider both abrupt transitions, which extend the abrupt $\Lambda_{\rm s}$CDM framework through sharp changes in the effective DE density behaviour, and smoother parametrisations that allow for a continuous evolution of the DE density.
These models have previously been investigated at both the background and perturbative levels at a theoretical level in Refs.~\cite{Bouhmadi-Lopez:2025ggl,Bouhmadi-Lopez:2025spo}.

\subsection{$\mathbf{{L\Lambda CDM}}$ \textit{model}}

The first of these extensions is the L$\Lambda$CDM model, a two-parameter extension of $\Lambda$CDM, where the DE density undergoes a transition described by a ladder-like function.
This model provides a simple generalisation of the abrupt $\Lambda_{\rm s}$CDM case, allowing the number of steps to be increased as desired. For numerical convenience, we will assume $N=8$ steps,

\begin{equation}
  \begin{aligned}
&  \Lambda_\mathrm{l}(z)=\Lambda \left[1-\frac{1}{10}\sum_{n=1}^{N} \mathcal{H}(z_n-z)\right],
\end{aligned}  
\label{densityladder}
\end{equation}
where $z_n=z_{f}+\frac{n}{8}(z_{i}-z_{f})$ denotes the redshift at which the transition between steps occurs, with $n \in [0, 8]$ representing the step index at that time, ordered from present to past, $\mathcal{H}(z_n - z)$ being the Heaviside step function evaluated at $z_n - z$ and $z_{i}$ and $z_{f}$ the redshift at which the transitions begins and ends respectively. In the case of an even number of steps, which we assume here, the initial and final transition redshifts are, respectively,
\begin{equation}
\begin{aligned}
    z_{i}=z_\dagger+\Delta z\cdot \frac{N}{2}, \\
    z_{f}=z_\dagger-\Delta z\cdot \frac{N}{2},
\end{aligned}
\end{equation}
where $\Delta z$ is the length of the steps, and $z_\dagger$ the redshift at which the DE density is 0. 

\subsection{$\mathbf{SSCDM}$ \textit{model}} 

The second of the extensions, another two parameter extension to $\Lambda$CDM, is the SSCDM model. In this models, the transition is governed by an interpolating function between two desired redshifts, $z_{i}$ and $z_{f}$. In this model, the DE density is assumed to be constant before and after the transitions:

\begin{equation}
  \begin{aligned}
&  \Lambda_\mathrm{ss}(x)=\Lambda \begin{cases}-1, & x \le x_{i},\\  1-2(126 t^5 - 420 t^6 \\
+ 540 t^7 - 315 t^8 + 70 t^9),  & x_{i}<x<x_{f},\\
1, & x\ge x_{f},\end{cases} \\
\end{aligned}  
\label{dnesitysscdm}
\end{equation}
where $t=\frac{x-x_{f}}{x_{i}-x_{f}}$ and $x=-\ln(1+z)$. We define $dz = z_i - z_f$ as a parameter that quantifies the redshift interval over which the transition occurs in this model.

 \subsection{$\mathbf{ECDM}$ \textit{model}}
 
 The last of the extensions is the ECDM model, where the DE density behaves as an error function\footnote{The error function is a function erf: $\mathbb{C}\rightarrow\mathbb{C}$ defined as: $$\operatorname{erf}(z)=\frac{2}{\sqrt{\pi}} \int_0^z e^{-t^2} d t\, .$$}, evolving in a continuous and uniform way for all redshifts. This interpolating function has two extra parameters when compared to $\Lambda$CDM,  $\eta$ the parameter that determines the smoothness of the transition, and $x_{\dagger} = -\ln(1 + z_{\dagger})$ that denotes the redshift at which the density changes sign:

\begin{equation}
    \Lambda_\mathrm{e}(x)=\Lambda \text{ Erf}[\eta(x-x_{\dagger})]/\text{ Erf}[-\eta x_{\dagger}].
    \label{densityecdm}
\end{equation}

For a detailed discussion of these smooth sign-switching extensions of abrupt $\Lambda_{\rm s}$CDM, we refer the reader to Refs.~\cite{Bouhmadi-Lopez:2025ggl,Bouhmadi-Lopez:2025spo}.

For convenience, we define a dimensionless energy density for the DE models introduced above as
\begin{equation}
f_{\textrm{d}}(z)=\Lambda_{\textrm{d}}(z)/\Lambda \, .
\label{def:f_d}
\end{equation}
Here, $\Lambda_{\textrm{d}}$ represents any of the DE energy densities given in Eqs.~(\ref{densityladder}), (\ref{dnesitysscdm}), and (\ref{densityecdm}).

\section{Observational Data and Methodology\label{sec3}}
\subsection{Data\label{sec3a}}

\begin{table}[t]
    \centering
    \renewcommand{\arraystretch}{1.4} 
    \begin{tabular}{clc}
\toprule \textbf{Param.} & \textbf{  Priors} & \textbf{Distribution}\\
\toprule 
$H_0$ & $\mathcal{U}(40,100)$ & $\mathcal{N}(69.0,5.0)$\\
$\Omega_{\textrm{m} 0}$ & $\mathcal{U}(0.05,0.99)$ & $\mathcal{N}(0.297, 0.05)$\\
$\Omega_{\textrm{b} 0}$ & $\mathcal{U}(0.01,0.10)$ & $\mathcal{N}(0.05,  0.005)$\\
\hline$z_{\dagger}$ & $\mathcal{U}(1.0,4.0)$ & flat\\
\hline$z_{\dagger}$ & $\mathcal{U}(1.0,4.0)$ & flat\\
$\Delta z$ & $\mathcal{U}(0.05,4.50)$ & flat\\
\hline$ z_{i}$ & $\mathcal{U}(1.1,4.5)$ & flat\\
$d z$ & $\mathcal{U}(0.1,4.5)$ & flat\\
\hline$z_{\dagger}$ & $\mathcal{U}(1.0,4.0)$ & flat\\
$\eta$ & $\mathcal{U}(0.1,30)$ & flat\\
\bottomrule
\end{tabular}
    \caption{\justifying{\textit{Priors adopted for the cosmological parameters used in the MCMC analysis. The quoted ranges correspond to the minimum and maximum values allowed for each parameter, as well as the initial distributions.}}}
    \label{tab:param}
\end{table}

We employ the Bayesian analysis framework Cobaya \cite{Torrado:2020dgo}, which uses a Markov Chain Monte Carlo (MCMC) sampler to generate the posterior distributions of the cosmological parameter space. The MCMC runs are performed in single-chain mode, and convergence is assessed using the default Gelman–Rubin $R-1$ statistic implemented in Cobaya, following the methodology described in \cite{Lewis:2013hha}. The analysis is based on the total likelihood function $\mathcal{L}_{\rm tot} \propto e^{-\chi^2_{\rm tot}/2}$, where the total chi-squared, $\chi^2_{\rm tot}$, is given by the sum of the contributions from the different cosmological data sets: $\chi^2_{\rm CMB}$, $\chi^2_{\rm BAO}$ and $\chi^2_{\rm SN}$.

We explore the parameter space described by the vector \footnote{We do not treat radiation as a free parameter. We derived it using $\Omega_{\rm m0}$ and $H_0$ through the following relations, $h\equiv H_0/100$,
$z_{eq} = 2.5\times 10^{4} \Omega_{\rm m0}  h^2 \left( \frac{T_{\rm CMB}}{2.7}\right)^{-4}$ and $ \Omega_{\rm r0}=\frac{\Omega_{\rm m0}}{1+z_{eq}} $\cite{Eisenstein:1997ik}. We set $T_{\rm CMB}=2.755$K \cite{Fixsen_2009}.}
\begin{equation}
\boldsymbol{\alpha} = \{ H_0, \Omega_{\rm m0}, \Omega_{\rm b0}, \boldsymbol{\Xi} \},
\end{equation}
where $H_0$ is the Hubble constant, $\Omega_{\rm m0}$ is the present-day matter density parameter, $\Omega_{\rm b0}$ is the baryon density parameter, and $\boldsymbol{\Xi}$ denotes the set of extension parameters, which depend on the specific model under consideration. In particular, for the abrupt $\Lambda_{\rm s}$CDM model we have $\boldsymbol{\Xi} = \{ z_{\dagger} \}$; for the L$\Lambda$CDM model, $\boldsymbol{\Xi} = \{ z_{\dagger}, \Delta z \}$; for the SSCDM model, $\boldsymbol{\Xi} = \{ z_i, dz \}$; and for the ECDM model, $\boldsymbol{\Xi} = \{ z_{\dagger}, \eta \}$.

A brief remark is warranted regarding the choice of prior range for the transition redshift $z_\dagger$. Previous analyses of abrupt sign-switching models have typically adopted priors restricted to $z_\dagger \in[1,3]$, motivated in part by early BAO measurements, including those from the Lyman-$\alpha$ forest. In the present work, we initially explored a similar range; however, we found that when including the full \textit{DESI DR2} data set, the posterior distribution of $z_\dagger$ is not fully contained within this range, indicating that such a restriction would artificially truncate the parameter space.

For this reason, we extend the prior range to allow $z_\dagger \in[1,4]$, ensuring that the posterior is fully captured and that the results are not prior-dominated at the boundary. We have verified that the main qualitative conclusions of the analysis remain stable under reasonable variations of the prior range. We further note that the preferred values of $z_\dagger$ lie above the redshift of the Lyman-$\alpha$ BAO measurement at $z_{\rm eff} \simeq 2.33$, and a full assessment of this apparent tension requires a perturbative treatment of structure formation, which is beyond the scope of the present background-level analysis and will be addressed in future work.

\subsubsection{Big Bang Nucleosynthesis (BBN) prior on $\Omega_{b0}h^2$}

In the absence of an external determination of the sound horizon at the drag epoch, $r_{\mathrm d}$, BAO measurements provide an uncalibrated standard ruler and therefore constrain the combination $H_0\, r_{\mathrm d}$, rather than $H_0$ and $r_{\mathrm d}$ separately. Although BAO data are also sensitive to $\Omega_{\mathrm m}$, an additional prior is required to fully break the degeneracy between the expansion rate and the sound-horizon scale.

We adopt a Gaussian prior on the physical baryon density, $\Omega_{\mathrm b0}h^2$, derived from Big Bang nucleosynthesis (BBN) considerations. 
Since our fundamental free parameter is $\Omega_{\mathrm b0}$ rather than $\Omega_{\mathrm b0}h^2$, this constraint is implemented through an additional likelihood term rather than as a direct prior. The inferred value of $\Omega_{\mathrm b0}h^2$ is obtained from observations of primordial light-element abundances and is largely insensitive to late-time cosmology. Specifically, we impose the constraint from \cite{Ong:2022wrsddff}

\begin{equation}
\Omega_{\mathrm b0}h^2 = 0.02233 \pm 0.00036.
\end{equation}

This enables calibrated BAO constraints without relying on the full CMB power spectrum. Our analysis is performed at the background level and assumes fixed neutrino properties, with no marginalisation over neutrino masses or relativistic degrees of freedom. This constraint is applied in all MCMC runs.

\subsubsection{Baryon Acoustic Oscillations}

BAO constitute a powerful observational probe for constraining cosmological parameters, particularly when combined with complementary datasets such as the CMB. The characteristic BAO feature imprinted in the matter power spectrum provides measurements of the Hubble expansion rate;  $H(z)$, and cosmological distance measures, thereby enabling constraints on the properties of DE through the late-time expansion history.

Under the assumption of standard pre-recombination physics, the comoving sound horizon at the baryon drag epoch, $r_{\mathrm d}$, is determined by the physical densities of baryons, cold dark matter, photons, and other relativistic species \cite{Brieden:2022heh}. In general, $r_{\mathrm d}$ can be expressed as a function of the physical baryon density $\omega_{\mathrm b} \equiv \Omega_{\mathrm b0}h^2$ and the total matter density $\omega_{\mathrm{bc}} \equiv (\Omega_{\mathrm b0} + \Omega_{\mathrm c0})h^2$, with a weaker dependence on the radiation content of the early Universe.

In this work, we do not compute $r_{\mathrm d}$ from first principles using a Boltzmann solver, nor do we vary parameters associated with the neutrino sector. Instead, we assume a standard early-time cosmology and calibrate the BAO scale through an external prior on $\Omega_{\mathrm b0}h^2$ derived from Big Bang nucleosynthesis. We adopt the fitting formula used on \cite{DESI:2025zgx}
\begin{equation}
r_d = 147.05~\mathrm{Mpc}
\left(\frac{\omega_{\mathrm b}}{0.02236}\right)^{-0.13}
\left(\frac{\omega_{\mathrm{bc}}}{0.1432}\right)^{-0.23},
\label{rddesi2}
\end{equation}
which is scaled to the best-fit values obtained from \textit{\textit{Planck 2018}} \cite{Planck:2018vyg}. Any residual dependence of $r_{\mathrm d}$ on relativistic species beyond photons is therefore implicitly fixed and not explored in this analysis.

Measurements of the BAO scale in the transverse and line-of-sight directions constrain complementary distance measures that depend only on the homogeneous expansion history. The transverse BAO signal probes the transverse comoving distance,
\begin{equation}
D_{\mathrm M}(z) = \frac{c}{H_0\sqrt{\Omega_{\mathrm k0}}}
\sinh\!\left[\sqrt{\Omega_{\mathrm k0}} \int_0^z \frac{\mathrm d z'}{E(z')}\right],
\end{equation}
where $\Omega_{\mathrm k0}$ denotes the present-day spatial curvature density parameter, $E(z) \equiv H(z)/H_0$ is the dimensionless Hubble expansion rate and $c$ is the speed of light. In the limit $|\Omega_{\mathrm k0}| \ll 1$, this reduces to the flat-universe expression
\begin{equation}
D_{\mathrm M}(z) = \frac{c}{H_0} \int_0^z \frac{\mathrm d z'}{E(z')}.
\end{equation}

The line-of-sight BAO signal constrains the Hubble expansion rate $H(z)$, or equivalently the radial distance
\begin{equation}
D_{\mathrm H}(z) = \frac{c}{H(z)}.
\end{equation}

Because BAO distances are measured relative to the comoving sound horizon at the drag epoch, the directly constrained observables are the dimensionless ratios $D_{\mathrm M}(z)/r_{\mathrm d}$ and $D_{\mathrm H}(z)/r_{\mathrm d}$. In addition, isotropic BAO measurements constrain the angle-averaged distance
\begin{equation}
D_{\mathrm V}(z) = \left[ z\,D_{\mathrm M}^2(z)\,D_{\mathrm H}(z) \right]^{1/3},
\end{equation}
again expressed relative to $r_{\mathrm d}$.

We utilise BAO measurements from the second data release of the Dark Energy Spectroscopic Instrument (DESI) \cite{DESI:2025zgx}, which combines observations of galaxies and quasars with Lyman-$\alpha$ forest tracers. The \textit{DESI DR2} BAO measurements span the redshift range $0.295 \leq z \leq 2.330$ and are provided in nine redshift bins, including both isotropic and anisotropic constraints. The data are expressed in terms of the transverse comoving distance $D_{\mathrm M}/r_{\mathrm d}$, the Hubble horizon $D_{\mathrm H}/r_{\mathrm d}$, and the angle-averaged distance $D_{\mathrm V}/r_{\mathrm d}$, together with their associated covariance matrices.

We define the goodness-of-fit to the BAO measurements using the standard chi-squared statistic,
\begin{equation}
\chi^2_{\mathrm{BAO}} = \boldsymbol{\Delta X}_{\mathrm{BAO}}^{\mathrm T}\,
\boldsymbol{\mathcal C}_{\mathrm{BAO}}^{-1}\,
\boldsymbol{\Delta X}_{\mathrm{BAO}},
\end{equation}
where $\boldsymbol{\Delta X}_{\mathrm{BAO}} = \mathbf{x}_{\mathrm{BAO}}^{\mathrm{obs}} - \mathbf{x}_{\mathrm{BAO}}^{\mathrm{th}}$ represents the difference between the observed BAO data vector and its theoretical prediction, and $\boldsymbol{\mathcal C}_{\mathrm{BAO}}^{-1}$ denotes the inverse of the BAO covariance matrix. Throughout this analysis, we make use of the full covariance matrix provided by the DESI collaboration, which contains non-vanishing off-diagonal elements arising from correlations between different redshift bins and BAO observables.

\subsection{Type Ia Supernovae}

Type Ia supernovae (SNIa) provide a key observational probe of the late-time expansion history of the Universe through their role as standardisable candles. In this work, we make use of the \textit{Pantheon+} $\&$ \textit{SH0ES} supernova compilations \cite{Brownsberger:2021uue} to constrain the cosmological parameters. The \textit{Pantheon+} sample comprises 1701 light curves corresponding to 1550 distinct SNIa, spanning the redshift range $0.001 \leq z \leq 2.26$. The \textit{SH0ES} calibration enters through the absolute magnitude normalisation of the low-redshift supernova sample and is consistently incorporated via the \textit{Pantheon+ $\&$ SH0ES} covariance matrix. Owing to their well-calibrated luminosities, these observations yield precise measurements of the distance modulus and are therefore highly sensitive to the background expansion history.

For each supernova, the observed distance modulus is provided directly in the compilation and is compared to the theoretical prediction
\begin{equation}
\mu_{\mathrm{SN}}^{\mathrm{th}}\left(z_{\mathrm{hel}}, z_{\mathrm{cmb}} \right)
= 5 \log_{10}\!\left[\frac{c}{H_0}D_L\left(z_{\mathrm{hel}}, z_{\mathrm{cmb}}\right)\right] + 25,
\end{equation}
where $z_{\mathrm{cmb}}$ and $z_{\mathrm{hel}}$ denote the redshifts measured in the CMB and heliocentric frames, respectively \cite{Conley:2011ku}. The Hubble-free luminosity distance is defined as
\begin{equation}
D_L\left(z_{\mathrm{hel}}, z_{\mathrm{cmb}} \right)
= \left(1+z_{\mathrm{hel}}\right)
\int_0^{z_{\mathrm{cmb}}}
\frac{\mathrm{d}z'}{E\left(z'\right)}.
\end{equation}

Since this analysis is restricted to background quantities, the theoretical prediction depends solely on the homogeneous expansion history.

The full covariance matrix $\boldsymbol{\mathcal C}_{\mathrm{SN}}$ provided with the \textit{Pantheon+ \& SH0ES} dataset incorporates statistical and systematic uncertainties, correlations among supernovae, and the Cepheid-based absolute magnitude calibration. Because the SH0ES distance ladder fixes the absolute luminosity scale, the nuisance parameter associated with the supernova absolute magnitude is already accounted for in the data release and is therefore not marginalised over in the likelihood analysis.

In practice, the likelihood treats the dataset as a combination of:
\begin{itemize}
    \item \textbf{Hubble-flow SN}, for which the theoretical distance modulus is evaluated from the cosmological model.
    \item \textbf{Cepheid-host (calibrator) SN}, for which the theoretical distance modulus is replaced by the measured Cepheid-based distances, $\mu_i^{\rm ceph}$.
\end{itemize}

Mathematically, this can be expressed as
\begin{widetext}
\begin{equation}
\mu_i^{\rm th} =
\begin{cases}
5 \log_{10}\big[(c/H_0) D_L(z_{\rm hel}, z_{\rm cmb})\big] + 25, & \text{Hubble-flow SN}, \\
\mu_i^{\rm ceph}, & \text{Cepheid-host SN}.
\end{cases}
\end{equation}
\end{widetext}

The residual vector is then defined as
\begin{equation}
\boldsymbol{\Delta\mu}_{\mathrm{SN}} \equiv \boldsymbol{\mu}_{\mathrm{SN}}^{\mathrm{obs}} - \boldsymbol{\mu}_{\mathrm{SN}}^{\mathrm{th}},
\end{equation}
and the corresponding chi-squared statistic is written in the standard Gaussian form
\begin{equation}
\chi_{\mathrm{SN}}^2
= \boldsymbol{\Delta\mu}_{\mathrm{SN}}^{\mathrm T}
\cdot \boldsymbol{\mathcal C}_{\mathrm{SN}}^{-1}
\cdot \boldsymbol{\Delta\mu}_{\mathrm{SN}},
\end{equation}
where the Hubble constant $H_0$ is fitted directly as a free parameter, reflecting the absolute magnitude calibration provided by the Cepheid distance ladder.

\subsubsection{Cosmic Microwave Background}

The CMB is one of the most powerful observational tools for studying the early Universe, carrying detailed
information about cosmological evolution from the epoch of photon decoupling to the present day. Although the full temperature and
polarisation power spectra depend on a wide range of physical effects, their sensitivity to the background expansion history can be accurately
summarised using a small number of geometric quantities. In particular, distance priors constructed from the angular size of the sound horizon
at decoupling retain much of the CMB’s constraining power on  DE and the background cosmology, while remaining applicable to a broad
range of cosmological models.

In this analysis, we employ a compressed CMB likelihood based on the \textit{Planck} 2018 data release \cite{Zhai:2018vmm}, which is
particularly well suited for studies focusing on background-level parameters. The likelihood is formulated in terms of the CMB shift
parameters—the acoustic scale $\ell_a$ and the shift parameter $R$—along with the present-day physical baryon density $\Omega_{\mathrm b0}h^2$.
These quantities capture the dominant geometric information encoded in the CMB anisotropy spectrum and allow CMB constraints to be incorporated
without explicitly computing the full set of angular power spectra. The shift parameters are defined as \cite{Wang:2007mza}

\begin{equation}
\begin{aligned}
&R \equiv \sqrt{\Omega_{\mathrm{m0}} H_0^2}\left(1+z_{\mathrm{CMB}}\right) \frac{D_A\left(z_{\mathrm{CMB}}\right)}{c}, \\
&l_a \equiv\left(1+z_{\mathrm{CMB}}\right) \frac{\pi D_A\left(z_{\mathrm{CMB}}\right)}{r_s\left(z_{\mathrm{CMB}}\right)},
\end{aligned}
\end{equation}
where $z_{\mathrm{CMB}}$ is the redshift at the decoupling epoch, $D_A\left(z_{\mathrm{CMB}}\right)$ is the angular diameter distance of photons in a flat FLRW universe expressed as

\begin{equation}
D_A(z)=\frac{1}{H_0(1+z)} \int_0^z \frac{d z^{\prime}}{E\left(z^{\prime}\right)}.
\end{equation}

The comoving sound horizon is given by

\begin{equation}
r_s(z)=\frac{c}{H_0} \int_z^\infty \frac{d z^{\prime}}{ E\left(z^{\prime}\right) \sqrt{3\left( 1+\bar{R}_b/(1+z^\prime)\right) }},
\label{cosounfhorizon}
\end{equation}
where $\bar{R}_b=31500 \Omega_{\mathrm{b}} h^2\left(T_{\mathrm{CMB}} / 2.7 K\right)^{-4}$, with $T_{\mathrm{CMB}}=2.7255 \mathrm{~K}$ \cite{Fixsen:2009ug}. The redshift at decoupling is given by the fitting formula \cite{Hu:1995en}

\begin{equation}
z_{\mathrm{CMB}}=1048\left[1+0.00124\left(\omega_{\mathrm{b}} \right)^{-0.738}\right]\left[1+g_1\left(\omega_{\mathrm{m}}\right)^{g_2}\right],
\end{equation}
where
\begin{equation}
\begin{aligned}
    & g_1=\frac{0.0783\left(\omega_{\mathrm{b}} \right)^{-0.238}}{1+39.5\left(\omega_{\mathrm{b}} \right)^{0.763}},  \\
    & g_2=\frac{0.56}{1+21.1\left(\omega_{\mathrm{b}} \right)^{1.81}}.
\end{aligned}
\end{equation}

The CMB covariance matrix is given by \cite{Zhai:2018vmm}

\begin{equation}
\boldsymbol{\mathcal{C}}_{\mathrm{CMB}}=10^{-\mathbf{8}} \times\left(\begin{array}{ccc}
1598.9554 & 17112.007 & -36.311179 \\
17112.007 & 811208.45 & -494.79813 \\
-36.311179 & -494.79813 & 2.1242182
\end{array}\right).
\end{equation}

Finally, the CMB contribution to the total $\chi^2$ is

\begin{equation}
\chi_{\mathrm{CMB}}^2=\boldsymbol{\Delta}\boldsymbol{X}_{\mathrm{CMB}}^{{T}} \cdot \boldsymbol{\mathcal{C}}_{\mathrm{CMB}}^{-1} \cdot \boldsymbol{\Delta}\mathbf{X}_{\mathrm{CMB}}\, ,
\end{equation}
where $\mathbf X_{\mathrm{CMB}}$ is the CMB parameters vector based on \textit{Planck 2018} release, as derived by \cite{Zhai:2018vmm}

\begin{equation}
\boldsymbol{\Delta}\boldsymbol{X}_{\mathrm{CMB}}=\left(\begin{array}{c}
R-1.74963 \\
l_a-301.80845 \\
\omega_{\mathrm{b}}-0.02237
\end{array}\right).
\end{equation}

\subsection{Information criteria\label{sec3b}}

We perform a statistical analysis to assess the performance of each sign-switching model relative to $\Lambda$CDM in fitting the observational data. To this end, we adopt an information-theoretic framework and focus on two of the most commonly used model-selection criteria in cosmology: the Akaike Information Criterion (AIC) \cite{Akaike:1974vps} and the Bayesian Information Criterion (BIC) \cite{Schwarz:1978tpv}. These statistics are straightforward to apply, as they depend only on the maximum likelihood attained by a given model, rather than on the full likelihood over the parameter space. We determine the best-fit parameters by maximising the likelihood using the minimizer implementation in Cobaya. In our analysis, we adopt the default BOBYQA algorithm \cite{cartis2018improvingflexibilityrobustnessmodelbased,Cartis_2021,Powell:2009NA06}, a derivative-free optimizer that is particularly well-suited for cosmological parameter estimation problems. Compared to standard gradient-based methods, BOBYQA generally exhibits improved robustness and efficiency when exploring complex likelihood surfaces with non-trivial parameter degeneracies. The resulting maximum-likelihood estimates are then used to compute information criteria for model comparison. This simplification, however, comes at the cost of relying on several underlying assumptions—most notably that the posterior distribution is Gaussian or approximately Gaussian—which may not hold accurately in practical applications \cite{Liddle:2007fy}. 

The AIC is defined in the following way
\begin{equation}
    {\rm AIC}=-2\ln{ \mathcal{L}_{\rm max}}+2\kappa,
\end{equation}
where $\mathcal{L}_{\rm max}$ is the maximum likelihood achieved by the model given the data and $\kappa$ the number of parameters of the model. This statistic provides a means of balancing goodness of fit against model complexity. The AIC is derived through an approximate minimisation of the Kullback--Leibler information entropy, which quantifies the discrepancy between the true data-generating distribution and that predicted by the model. For a detailed statistical justification, see \cite{2002}. Nevertheless, when dealing with small data samples, it is more appropriate to employ the corrected Akaike Information Criterion, AIC$_{\mathrm c}$ \cite{Sugiura:1978vps}

\begin{equation}
    {\rm AIC}_{\mathrm c}={\rm AIC}+
\frac{2\kappa^2 + 2\kappa
}{M-\kappa-1},
\end{equation}
where $M$ denotes the total number of data
points used in the fit. For $M\gg\kappa$, ${\rm AIC}_{\rm c}\simeq{\rm AIC}$. To compare the effectiveness of each models against $\Lambda$CDM we compare the difference between both models, $\Delta{\rm AIC}_{\rm c}={\rm AIC}_{\rm c, model}-{\rm AIC}_{\rm c, \Lambda CDM}$. In this case, due to the different number of free parameters in each models, the selection of the best one is not straightforward, as the preferred one need not be  the one with the lowest $\chi^2$ due to the penalisation for the extra parameters. To interpret the $\Delta$AIC results, we follow Jeffreys’s scale \cite{Jeffreys:1939xee}, summarised in Tab. \ref{deltaaic}.

The BIC incorporates a different penalisation for model complexity, 
\begin{equation}
    {\rm BIC}=-2\ln{ \mathcal{L}_{\rm max}}+2\kappa\ln{M}.
\end{equation}

While the AIC tends to favour models that improve the fit regardless of the added complexity, the BIC imposes a stronger penalty for additional parameters, an effect that becomes particularly significant for large data sets. This renders the BIC a more conservative criterion for evaluating model preference. The BIC rests on the assumption that the data points are independent and identically distributed, an assumption that may not hold for all data sets under consideration.

\begin{table}
\centering
\renewcommand{\arraystretch}{1.4} 
\begin{tabular}{ccl}
\toprule$\mathbf{\Delta}$\bf AIC/$\mathbf{\Delta}$\bf BIC & & \bf Interpretation \\
\toprule $>10$ & & Desively disfavoured \\
\hline $5 \sim 10$ & & Strongly disfavoured \\
\hline $2 \sim 5$ & & Moderately disfavoured \\
\hline$-2 \sim 2$ & & Compatible \\
\hline$-5 \sim-2$ & & Moderately favoured \\
\hline$-10 \sim-5$ & & Strongly favoured \\
\hline$<-10$ & & Decisively favoured \\
\bottomrule
\end{tabular}
\caption{\justifying{\textit{The differences in the Akaike and Bayesian information criteria, $\Delta \mathrm{AIC}$ and $\Delta \mathrm{BIC}$, may be interpreted according to Jeffreys's scale. This scale provides a systematic, qualitative framework for assessing the relative performance of models with respect to a reference model, which in the present work is $\Lambda \mathrm{CDM}$. Negative values of $\Delta \mathrm{AIC}$ or $\Delta \mathrm{BIC}$ indicate that the alternative model is favoured, whereas positive values signify a preference for the $\Lambda \mathrm{CDM}$ scenario.
}}}
\label{deltaaic}
\end{table}

In practice, the AIC and BIC often yield broadly consistent qualitative conclusions, although differences may arise in the detailed ranking of competing models. When applying these criteria, it is essential to assess how well the assumptions underlying their derivation are satisfied in realistic analyses. A particularly relevant issue is the presence of parameter degeneracies: parameters that are weakly constrained or effectively unconstrained are penalised by both the AIC and BIC, whereas they do not influence the Bayesian evidence in the same manner. Consequently, interpreting the BIC as a reliable estimator of evidence differences becomes questionable in such situations.

\section{Observational constraints}\label{sec3c}

We now present the observational constraints on the four sign-switching DE models and compare them with the standard $\Lambda$CDM scenario. The discussion is organised into three subsections.

We first analyse the inferred parameter values and their posterior distributions for the following data-set combinations:
\begin{itemize}
    \item \textbf{Combination I:} cosmic microwave background (CMB)
    \item \textbf{Combination II:} baryon acoustic oscillations (BAO)
    \item \textbf{Combination III:} type I supernovae (SN)
    \item \textbf{Combination IV:} cosmic microwave background and baryon acoustic oscillations (CMB+BAO).
    \item \textbf{Combination V:} cosmic microwave background, baryon acoustic oscillations and type I supernovae (CMB+BAO+SN).
\end{itemize}
Second, we evaluate the relative statistical performance of the models using information criteria applied to the same data combinations. Finally, we reconstruct selected background and cosmographic quantities.

\subsection{Mean values and standard deviations of cosmological parameters.}

\begin{table*}[htbp]
    \centering
    \renewcommand{\arraystretch}{1.4} 
    \begin{tabular}{lcccccc
}
\hline\hline \textbf{Model} & {\bf H$\mathbf{_0}$} & {$\mathbf{\Omega}_{\rm \textbf{m0}}$} & $\mathbf{10^2}${$\mathbf{\Omega_{\rm \textbf{b} \mathbf{0}}h^2}$} & {\bf z$_\mathbf{\dagger}$\bf /z$_\mathbf{i}$} & $\boldsymbol{\Delta z/}{\bf dz/}\boldsymbol{\eta}$  \\
\hline\hline
 \multicolumn{6}{c}{\textbf{CMB}} \\
 $\Lambda$CDM          & $67.33\pm0.54$ & $0.3156\pm0.0074$ & $2.2362\pm0.0141$ &                   &  \\
 $\Lambda_{\mathrm{s}}\mathrm{CDM}$    & $70.06^{+0.25}_{-2.80}$ & $0.2928^{+0.0251}_{-0.0059}$  & $2.2362\pm0.0135$ & $>1.35 (95\% $CL)  & \\
 L$\Lambda$CDM         & $71.40^{+0.69}_{-4.02}$ & $0.2829^{+0.0329}_{-0.0101}$ & $2.2361\pm0.0134$ & $>1.58 (95\% $CL)  & $<0.79 (95\% $CL)  \\
 SSCDM                 & $71.32^{+0.53}_{-4.08}$ & $0.2837^{+0.0336}_{-0.0090}$ & $2.2359\pm0.0135$ & $>1.84 (95\% $CL) & $<3.60 (95\% $CL)  \\
 ECDM                  & $70.31^{+0.34}_{-3.11}$ & $0.2909^{+0.0272}_{-0.0676}$ & $2.2359\pm0.0136$ & $>1.35 (95\% $CL) & unconstrained \\
 \hline
  \multicolumn{6}{c}{\textbf{BAO}}  \\
 $\Lambda$CDM          & $68.75\pm0.45$ & $0.2973\pm0.0087$  & $2.2347\pm0.0364$ &                   &\\
 $\Lambda_{\mathrm{s}}\mathrm{CDM}$     & $68.72\pm0.43$     & $0.2975\pm0.0087$ & $2.2322\pm0.0357$ & $>2.41 (95\% $CL)  &  \\
 L$\Lambda$CDM         & $69.66\pm0.78$ & $0.3224\pm0.0184$  & $2.2326\pm0.0357$ & $>1.56 (95\% $CL) & $0.36^{+0.13}_{-0.22}$ \\
 SSCDM                 & $69.58\pm1.04$ & $0.3214\pm0.0265$  & $2.2322\pm0.0361$ & $>2.77 (95\% $CL) & $<4.11 (95\% $CL) \\
 ECDM                  & $68.94^{+0.37}_{-0.69}$ & $0.3035^{+0.0066}_{-0.0155}$  & $2.2329\pm0.0356$ & $>1.63 (95\% $CL) & $>2.23 (95\% $CL) \\
 \hline
  \multicolumn{6}{c}{\textbf{SN}} \\
 $\Lambda$CDM          & $73.49\pm1.02$ & $0.3359\pm0.0190$ & $2.2330\pm0.0362$ &               &  \\
 $\Lambda_{\mathrm{s}}\mathrm{CDM}$       & $73.52\pm1.01$  & $0.3336\pm0.0180$ & $2.2311\pm0.0360$ & unconstrained &  \\
 L$\Lambda$CDM         & $73.51\pm1.02$ & $0.3385\pm0.0251$ & $2.2325\pm0.0351$ & unconstrained & $<0.72 (95\% $CL) \\
 SSCDM                 & $73.49\pm1.02$ & $0.3359^{+0.018}_{0.020}$ & $2.2324\pm0.0357$  & unconstrained & $<3.58 (95\% $CL) \\
 ECDM                  & $73.50\pm1.00$ & $0.3347^{+0.0177}_{-0.0198}$  & $2.2333\pm0.0359$ & unconstrained & unconstrained \\
 \hline
 \multicolumn{6}{c}{\textbf{CMB + BAO}}\\
 $\Lambda$CDM          & $68.41\pm0.29$ & $0.3009\pm0.0037$ & $2.2516\pm0.0122$ &                   &   \\
 $\Lambda_{\mathrm{s}}\mathrm{CDM}$        & $68.75\pm0.33$ & $0.3013\pm0.0038$ & $2.2402\pm0.0124$ & $3.01^{+0.26}_{-0.60}$     &   \\
 L$\Lambda$CDM         & $68.79^{+0.31}_{-0.35}$ & $0.3014\pm0.0038$ & $2.2387\pm0.0128$ & $>2.66 (95\% $CL) & $<0.64 (95\% $CL) \\
 SSCDM                 & $68.76\pm0.32$ & $0.3014\pm0.0038$ & $2.2399\pm0.0125$ & $>2.79 (95\% $CL) & $<2.66 (95\% $CL) \\
 ECDM                  & $68.74\pm0.32$ & $0.3015\pm0.0038$ & $2.2401\pm0.0126$ & $3.06^{+0.27}_{-0.60}$     & $>3.40 (95\% $CL)\\
 \hline
 \multicolumn{6}{c}{\textbf{CMB + BAO + SN}} \\
 $\Lambda$CDM          & $68.73\pm0.28$ & $0.2972\pm0.0035$ & $2.2607\pm0.0119$ &                   &  \\
 $\Lambda_{\mathrm{s}}\mathrm{CDM}$        & $69.12\pm0.31$ & $0.2980\pm0.0035$ & $2.2455\pm0.0124$ & $2.77^{+0.10}_{-0.44}$ &  \\
 L$\Lambda$CDM         & $69.28^{+0.33}_{-0.43}$ & $0.2977\pm0.0037$ & $2.2406\pm0.0135$ & $>2.59 (95\% $CL) &  $<0.737 (95\% $CL) \\
 SSCDM                 & $69.12\pm0.30$ & $0.2980\pm0.0035$ & $2.2455\pm0.0123$ & $3.56^{+0.70}_{-0.52}$ & $<2.80 (95\% $CL) \\
 ECDM                  & $69.14^{+0.30}_{-0.64}$ & $0.2979\pm0.0036$ & $2.2450\pm0.0126$ & $2.86^{+0.16}_{-0.49}$     &  $>1.96 (95\% $CL) \\ 
 \hline\hline
\end{tabular}
    \caption{ \justifying{\textit{Mean values and standard deviations of the cosmological parameters obtained for each sign-switching models, and for $\Lambda$CDM paradigm, under the five different dataset combinations considered in this work: \textbf{Combination I} (CMB), \textbf{Combination II} (BAO), \textbf{Combination III} (SN), \textbf{Combination IV} (CMB+BAO)  and \textbf{Combination V} (CMB+BAO+SN). }}}
    \label{tab:obscons}
\end{table*}

\begin{figure*}[t]
\centering

\begin{subfigure}{0.4\textwidth}
\centering
\includegraphics[width=\textwidth]{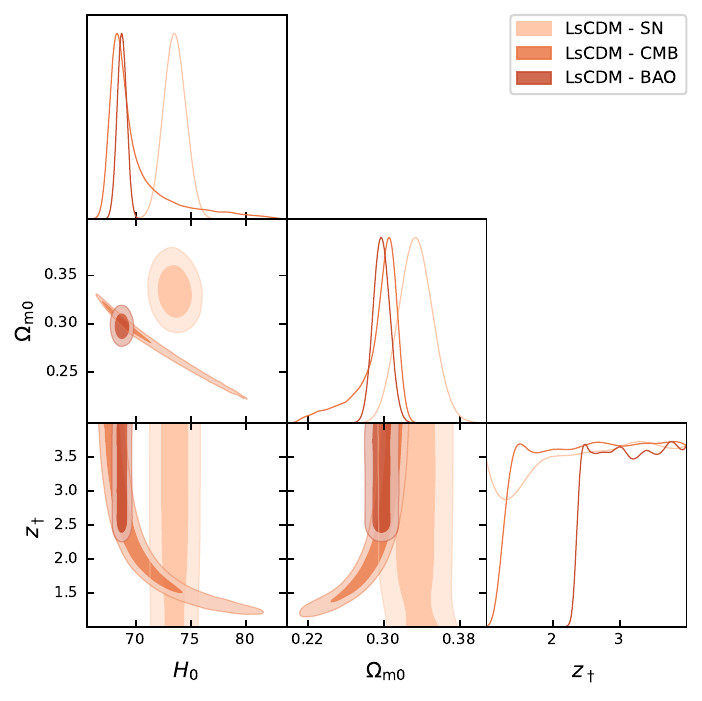}
\caption*{(A) Abrupt $\Lambda_{\rm s}$CDM}
\end{subfigure}
\hspace{0.05\textwidth}
\begin{subfigure}{0.4\textwidth}
\centering
\includegraphics[width=\textwidth]{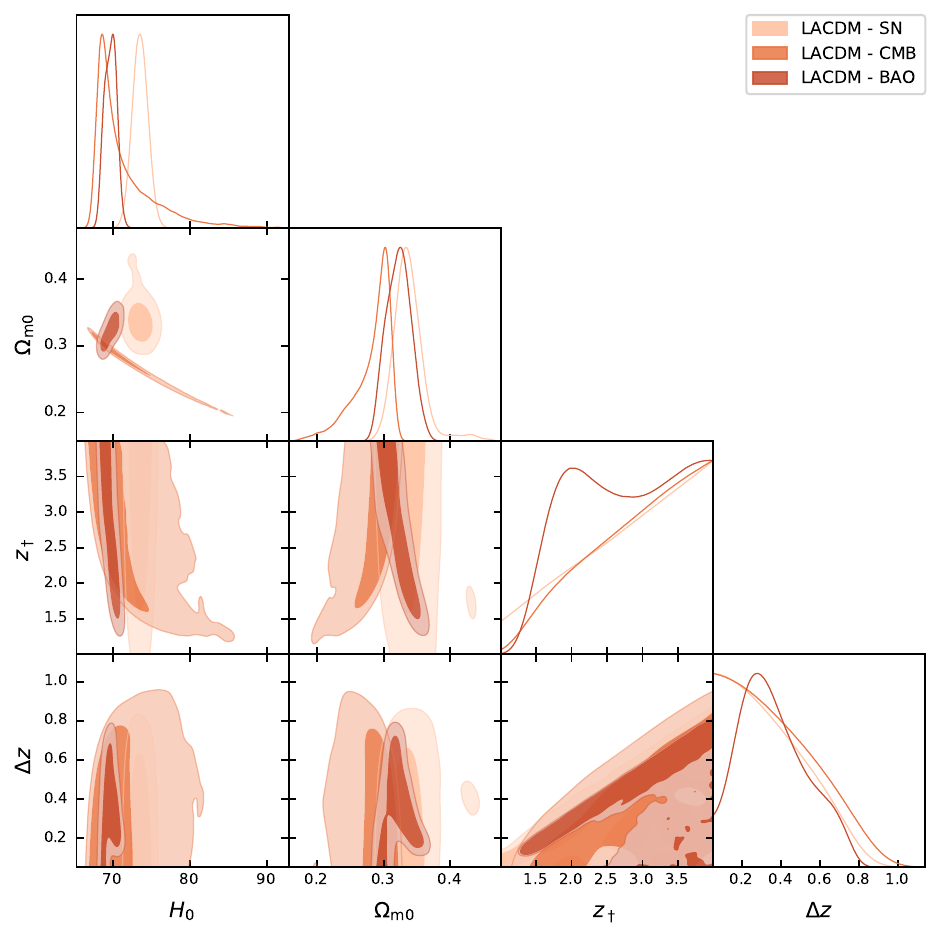}
\caption*{(B) L$\Lambda$CDM}
\end{subfigure}

\vspace{0.5cm}

\begin{subfigure}{0.4\textwidth}
\centering
\includegraphics[width=\textwidth]{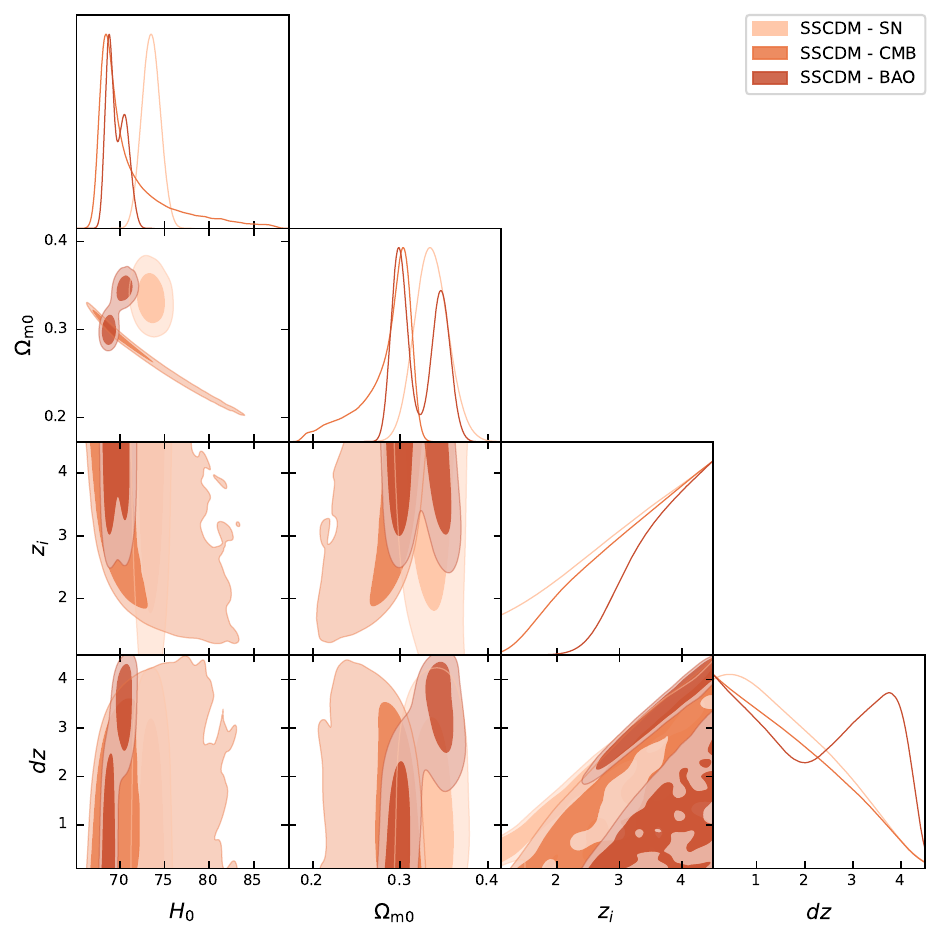}
\caption*{(C) SSCDM}
\end{subfigure}
\hspace{0.05\textwidth}
\begin{subfigure}{0.4\textwidth}
\centering
\includegraphics[width=\textwidth]{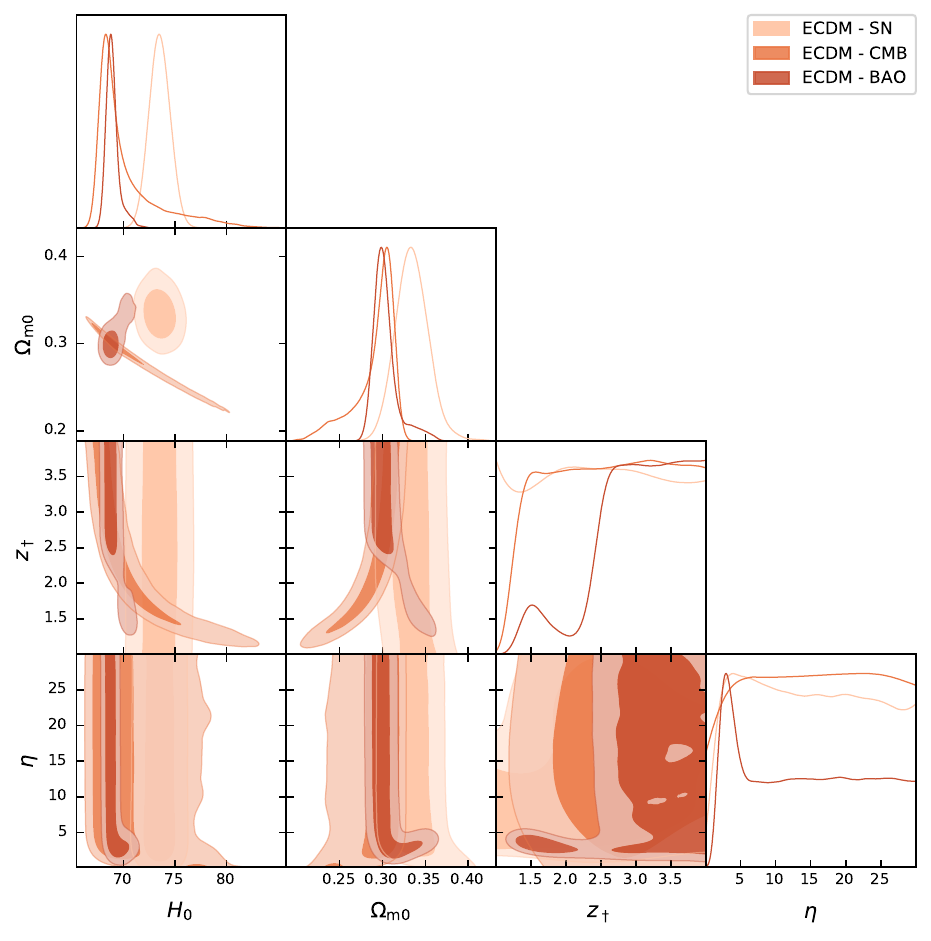}
\caption*{(D) ECDM}
\end{subfigure}

\caption[BAO comp1]{\justifying{\textit{Two-dimensional posterior distributions for the
sign-switching models, using the latest \textit{DESI DR2} (BAO) release,
Planck18 (CMB) and Pantheon+ \& SH0ES (SN) datasets. The contours correspond to the
$68\%$ and $95\%$ confidence levels (C.L.).}}}
\label{fig:cmb_bao_sn}

\end{figure*}

In this subsection, we constrain the cosmological parameters and report their marginalised means and standard deviations. Specifically, we determine the aforementioned free parameters: the Hubble parameter $H_0$, the present matter density $\Omega_{\rm m0}$, the present baryon density $\Omega_{\rm b0}$, the transition (onset) redshifts $z_\dagger/z_i$, and the transition rate for the smooth extensions, $\Delta z/dz/\eta$. The results are summarised in Table~\ref{tab:obscons}; see also Fig.~\ref{fig:cmb_bao_sn} and ~\ref{fig:cmb+bao+sn}.

\subsubsection{CMB constraints}

Combination I consists solely of CMB data. Fitting the models to these data reveals a significant shift in the inferred value of $H_0$ relative to $\Lambda$CDM. In particular, sign-switching models favour higher values of $H_0$, especially in their smooth realisations.

The discontinuous models, abrupt $\Lambda_{\rm s}$CDM and L$\Lambda$CDM, favour higher inferred values of the Hubble parameter, $H_0 \sim 70\text{--}71$ km s$^{-1}$ Mpc$^{-1}$. This shift is primarily driven by an extended right-hand tail in the posterior, which pulls the mean towards larger values. The continuous models, SSCDM and ECDM, show a comparable improvement, most notably for SSCDM, where the higher mean value of $H_0$ is accompanied by an increased standard deviation. ECDM yields constraints similar to those of abrupt $\Lambda_{\rm s}$CDM. These results are consistent with the \textit{Riess} determination of $H_0 = 73.03 \pm 1.42$ km s$^{-1}$ Mpc$^{-1}$ and the new \textit{H0DN} consensus $H_0=73.50 \pm 0.81$ km s$^{-1}$ Mpc$^{-1}$ \cite{H0DN:2025lyy}; however, this consistency is largely driven by the broader uncertainty relative to $\Lambda$CDM.

This combination also favours a lower matter density than $\Lambda$CDM, with the higher inferred value of $H_0$ implying a larger present-day DE component.

Regarding the additional model parameters, although they remain weakly constrained, all scenarios favour a sign-switching redshift of $z_\dagger \gtrsim 1.4$ at the 95\% confidence level. Furthermore, the transition is inferred to span a redshift interval of roughly $\lesssim 3.5\text{--}4.5$, pointing to a gradual evolution rather than a sharp transition, in contrast with earlier expectations.

It is worth noting that our constraints on the $\Lambda_{\mathrm{s}}\mathrm{CDM}$ model differ slightly from those reported in 
\cite{Akarsu:2021fol,Akarsu:2022typ,Akarsu:2023mfb,Escamilla:2025imi}. These differences primarily stem from our adoption of broader priors on the sign-switching redshift, $z_\dagger$. Specifically, by adopting a uniform prior $\mathcal{U}[1.0,4.0]$ for this parameter, we obtain a lower inferred mean value of $H_0$ and a correspondingly higher present-day matter density parameter, $\Omega_{\rm m0}$. This behaviour reflects the fact that the interval $3 \lesssim z_\dagger \lesssim 4$ is associated with lower values of $H_0$ and higher values of $\Omega_{\rm m0}$. When the prior range is instead restricted to $\mathcal{U}[1.0,3.0]$, as in \cite{Akarsu:2022typ,Akarsu:2023mfb,Escamilla:2025imi}, we recover parameter estimates of $H_0 = 70.87 \pm 2.73~\mathrm{km\,s^{-1}\,Mpc^{-1}}$ and $\Omega_{\rm m0} = 0.2861 \pm 0.0212$.

A comment is in order regarding our use of the compressed \textit{Planck 2018} likelihood based on CMB shift parameters. In this work, we restrict the analysis to background-level observables and do not assume a specific perturbative realisation of the sign-switching dark energy models. Within this context, the shift-parameter approach provides an effective summary of the geometric information encoded in the CMB. We have explicitly verified that, for $\Lambda$CDM and for the abrupt $\Lambda_{\rm s}$CDM scenario, this compressed likelihood reproduces results consistent with those obtained using the full CMB likelihood at the background level.

Nevertheless, this approach entails a loss of constraining power compared to the full CMB analysis, particularly for parameters correlated with $H_0$ and for models exhibiting non-trivial late-time deviations from $\Lambda$CDM. As a consequence, parameters governing the transition, such as the transition redshift $z_\dagger$ and the transition rate, remain only weakly constrained. Similarly, any shifts in the inferred value of $H_0$ should be interpreted with appropriate caution. The results presented here should therefore be regarded as a first assessment of the background-level viability of these models, while a full analysis including perturbations and the complete CMB likelihood is left for future work.

\subsubsection{BAO constraints}

Data Combination II probes the evolution of the BAO measurements. The parameters $H_0$ and $\Omega_{\rm m0}$ appear to increase across all models relative to $\Lambda$CDM, particularly in L$\Lambda$CDM and SSCDM.

Regarding the additional parameters, the models exhibit mildly different behaviour. For abrupt $\Lambda_{\rm s}$CDM, the transition redshift lies beyond the range currently probed by DESI: although the highest effective redshift is $z_{\rm eff}=2.33$, we obtain a lower bound of $z_\dagger > 2.41$ at the 95\% CL. This indicates that, within abrupt $\Lambda_{\rm s}$CDM, the DESI data disfavour a negative DE density. Nevertheless, the smooth sign-switching models exhibit somewhat different behaviour. Previous studies of this class of models typically favoured rapid transitions \cite{Akarsu:2019hmw,Akarsu:2025gwi} in order to mimic the behaviour of abrupt $\Lambda_{\rm s}$CDM. In contrast, we find that these models favour slower transitions (cf. Fig.~\ref{fig:cmb_bao_sn}):
\begin{itemize}
    \item In L$\Lambda$CDM, inspection of the posterior distribution shown in Fig.~\ref{fig:cmb_bao_sn} indicates a preferred step length of $\Delta z_{\rm step}\sim0.3$, corresponding to a transition spanning $\sim2.4$ in redshift and therefore affecting late-time dynamics. 
    \item The SSCDM model exhibits a bimodal distribution, where a preferred transition length of $\sim3.5$ in redshift appears to allow for even higher values of $H_0$ and $\Omega_{\rm m0}$. 
    \item The ECDM model likewise favours a smaller transition speed, $\eta\sim4$, corresponding to a transition spanning approximately $3.5$ in redshift. 
\end{itemize}
These results suggest that, although the transition redshift is typically pushed beyond the largest effective redshift of the dataset, slow sign-switching is associated with a smaller DE density prior to the present epoch.

\begin{figure*}[t]
\centering

\begin{subfigure}{0.4\textwidth}
\centering
\includegraphics[width=\textwidth]{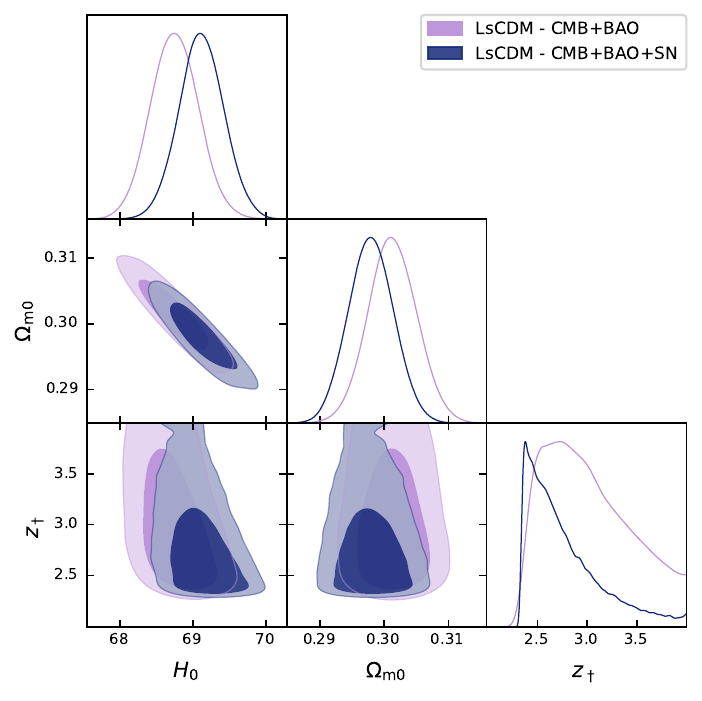}
\caption*{(A) Abrupt $\Lambda_{\rm s}$CDM}
\end{subfigure}
\hspace{0.05\textwidth}
\begin{subfigure}{0.4\textwidth}
\centering
\includegraphics[width=\textwidth]{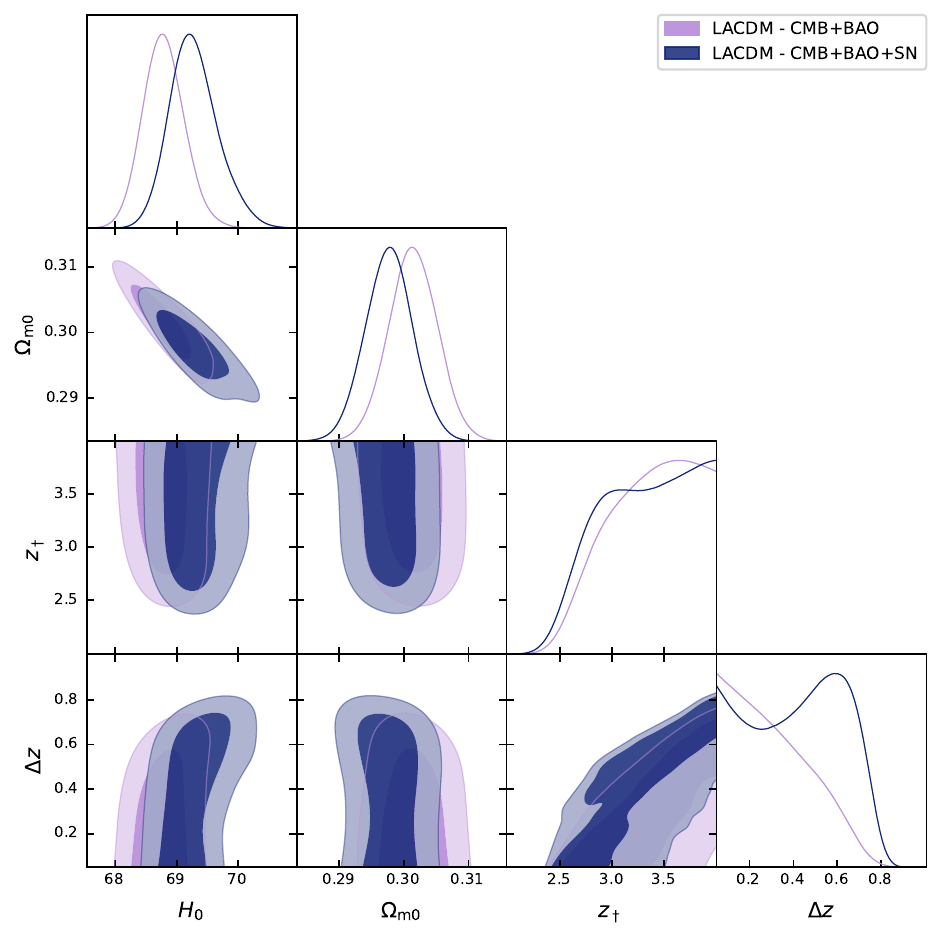}
\caption*{(B) L$\Lambda$CDM}
\end{subfigure}

\vspace{0.5cm}

\begin{subfigure}{0.4\textwidth}
\centering
\includegraphics[width=\textwidth]{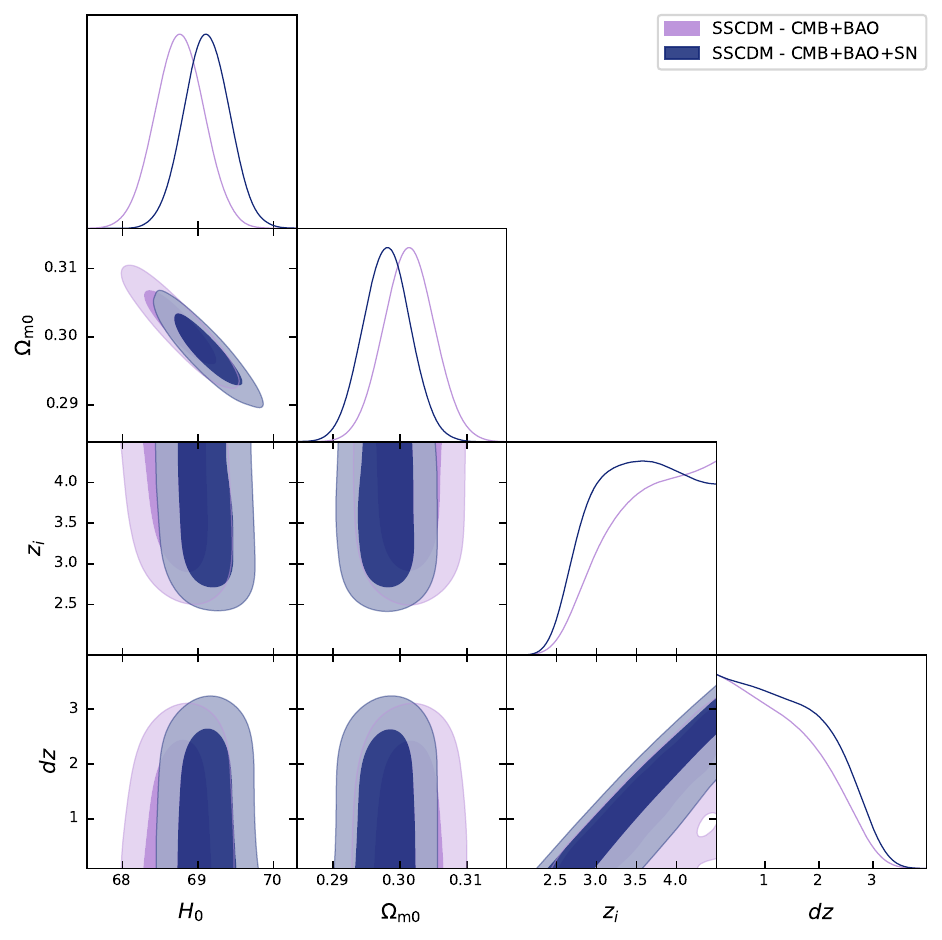}
\caption*{(C) SSCDM}
\end{subfigure}
\hspace{0.05\textwidth}
\begin{subfigure}{0.4\textwidth}
\centering
\includegraphics[width=\textwidth]{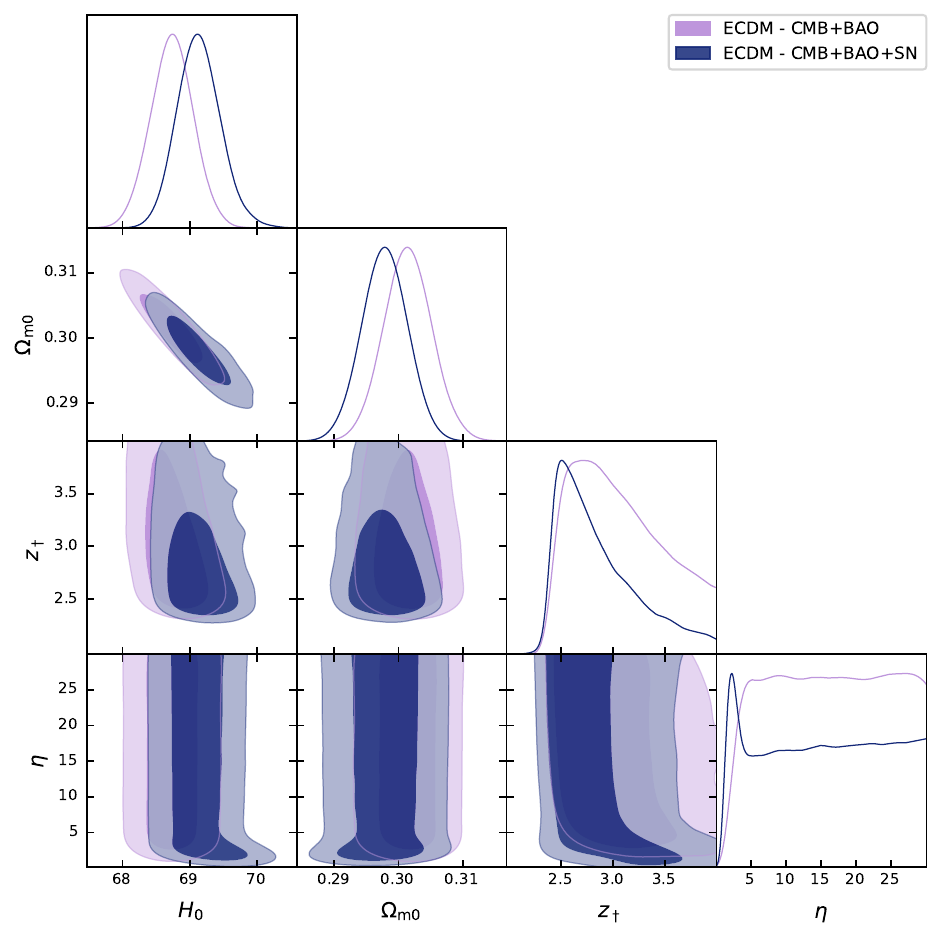}
\caption*{(D) ECDM}
\end{subfigure}

\caption[BAO comp2]{\justifying{\textit{Two-dimensional posterior distributions for the
sign-switching models, using the combinations \textit{DESI DR2} (BAO) +
Planck18 (CMB) and  \textit{DESI DR2} (BAO) +
Planck18 (CMB) + Pantheon+ $\&$ SH0ES (SN) datasets. The contours correspond to the
$68\%$ and $95\%$ confidence levels (C.L.).}}}
\label{fig:cmb+bao+sn}

\end{figure*}

\subsubsection{SN constraints}

Data Combination III relies only on SN observations. When fitted to these data, both the Hubble constant $H_0$ and the matter density $\Omega_{\rm m0}$ are similarly constrained across all models, exhibiting nearly identical central values and uncertainties. This dataset alone has limited constraining power on the additional parameters, failing to constrain $z_\dagger/z_i$ and generally favouring rapid transitions.

\subsubsection{CMB and BAO constraints}

Data Combination IV includes the CMB and BAO observations introduced in the previous section. This combination provides stronger constraints on the sign-switching redshift.

For the Hubble constant $H_0$ and the matter density $\Omega_{\rm m0}$, all sign-switching models favour slightly higher values than those inferred under $\Lambda$CDM. By contrast, the baryon density $\Omega_{\rm b0}h^2$ is inferred to be slightly lower. Relative to previous analyses of the $\Lambda_{\mathrm{s}}\mathrm{CDM}$ model, the adoption of a broader prior on the sign-switching redshift $z_\dagger$ leads to higher preferred values than those reported in earlier studies \cite{Akarsu:2021fol,Akarsu:2022typ,Akarsu:2023mfb,Escamilla:2025imi}.

Regarding the smoother sign-switching models: (i) the L$\Lambda$CDM model also exhibits higher values of $z_\dagger$, although it remains unconstrained from above; similarly, the transition speed is not constrained and favours rapid transitions. (ii) The SSCDM model shows behaviour similar to that of L$\Lambda$CDM, with higher values of $z_\dagger$ and lower values of $\Delta z$ compared to previous combinations. (iii) The ECDM model constrains the transition redshift $z_\dagger$ to values comparable to those of $\Lambda_{\mathrm{s}}\mathrm{CDM}$ and does not exhibit a preferred transition speed, provided the transition is rapid, thereby closely mimicking the $\Lambda_{\mathrm{s}}\mathrm{CDM}$ behaviour in this case.

\subsubsection{CMB, BAO and SN constraints}

Data Combination V incorporates all previously considered datasets, namely CMB, BAO, and SN. When fitted to this full dataset, all sign-switching models favour higher inferred values of $H_0$ and $\Omega_{\rm m0}$ than $\Lambda$CDM, while the baryon density is inferred to be lower, reflecting the impact of the sign-switching behaviour.

The transition redshift is constrained for both $\Lambda_{\mathrm{s}}\mathrm{CDM}$ and ECDM, yielding a mean value of $z_\dagger \sim 2.80$. In the SSCDM model, the onset redshift exhibits a preferred value near $z_\dagger \sim 3.5$; however, the absence of an upper bound indicates that the parameter remains weakly constrained. Only a lower bound on the transition redshift can be established in L$\Lambda$CDM, reflecting the limited sensitivity of the current data at these redshifts.

Turning to the transition speed, although the parameter is not tightly constrained, the posterior distributions shown in Fig.~\ref{fig:cmb+bao+sn} indicate that L$\Lambda$CDM favours a value of $\Delta z \sim 0.6$, corresponding to a transition length of $N\cdot\Delta z \sim 4.8$. For ECDM, the transition speed parameter $\eta$ favours $\eta \sim 3.5$, implying a transition spanning approximately $4$ in redshift. Both cases therefore correspond to relatively slow transitions and represent a noticeable departure from the abrupt behaviour characteristic of the $\Lambda_{\mathrm{s}}\mathrm{CDM}$ model. 

Nevertheless, the data remains compatible with fast transitions, and both regimes should be considered. This is illustrated by the SSCDM model, in which faster transitions are preferred. In principle, slower transitions could provide a greater alleviation of the Hubble tension, particularly if the transition were to occur at lower redshifts than those favoured by our analysis. However, the preference for sign-switching at higher redshifts found here appears to be primarily driven by the \textit{DESI DR2} dataset, which disfavors a negative DE density between its data points.

This behaviour may be related to the partial dependence of three-dimensional BAO measurements on the $\Lambda$CDM model. Should transverse BAO data from future DESI releases be incorporated, these conclusions could change significantly, as such measurements appear to accommodate lower sign-switching redshifts \cite{Akarsu:2021fol,Akarsu:2022typ,Akarsu:2023mfb,Akarsu:2024eoo}. Finally, we note that our values for $\Lambda_{\mathrm{s}}\mathrm{CDM}$ differ slightly from those obtained for Combinations IV and V in \cite{Yadav:2025vpx}, primarily due to the broader prior adopted for $z_\dagger$.

\subsubsection{Further discussions}

Exploring alternative prior choices reveals that, in some cases, the width of the prior significantly impacts the resulting posterior distribution. For the parameter $\eta$ in the ECDM model, extending the upper bound of the prior beyond $\mathcal{U}(0.1,30)$ removes the preference for slower transition speeds, instead producing a plateau for $\eta \gtrsim 3.3$. In principle, sufficiently rapid transitions should not exhibit a preferred value, as the current datasets are not precise enough to constrain this regime.

This behaviour highlights the strong sensitivity of the inferred constraints to prior assumptions, a trend that is also observed across the other models. These findings motivate a reassessment of the commonly adopted assumption that transitions must be fast, an idea inherited from graduated dark energy models but not necessarily required by the underlying physics. Relaxing this assumption opens a broader phenomenological landscape in which slower transitions may play a significant role.

In the specific case of the SSCDM model, the inferred constraints depend sensitively on the chosen parametrisation. Treating the initial redshift $z_i$ as a free parameter enables the transition speed to be constrained via an upper bound on the transition width. By contrast, parametrising the model in terms of the final redshift $z_f$ does not yield comparable constraints, despite its more direct connection to the present-day DE density. This behaviour highlights the strong prior dependence of the model and suggests that physically motivated consistency conditions, such as enforcing $z_f > 0$, may be required to obtain meaningful constraints.

\subsection{Statistical Model Selection}

\begin{table}[t]
    \centering
    \renewcommand{\arraystretch}{1.4} 
    \begin{tabular}{lcccccc
}
\hline\hline
\textbf{Model} & $\chi^2 $& {$\Delta$AIC} & {$\Delta$AIC$_{\rm c}$} & {$\Delta$BIC} \\
\hline\hline
\multicolumn{5}{c}{\textbf{SN}} \\
$\Lambda$CDM          & 1452.017 & 0.000 & 0.000 & 0.000 \\
$\Lambda_{\mathrm{s}}\mathrm{CDM}$       & 1451.885 & 1.868 & 1.643 & 7.307 \\
L$\Lambda$CDM         & 1451.084 & 2.436 & 2.457 & 13.315  \\
SSCDM                 & 1449.660 & 1.643 & 1.664 & 12.522 \\
ECDM                  & 1451.081 & 3.064 & 3.086 & 13.943 \\
\hline
\multicolumn{5}{c}{\textbf{BAO}} \\
$\Lambda$CDM          & 10.283 & 0.000 & 0.000 & 0.000\\
$\Lambda_{\mathrm{s}}\mathrm{CDM}$       & 10.284 & 2.000 & 4.044 & 2.639\\
L$\Lambda$CDM         & 4.632 & -1.652 & 3.448 & -0.374 \\
SSCDM                 & 6.509 &  0.225 & 5.325 & 1.503 \\
ECDM                  & 6.896 &  0.613 & 5.713 & 1.891 \\
\hline
\multicolumn{5}{c}{\textbf{CMB+BAO}} \\
$\Lambda$CDM          & 16.244 & 0.000 & 0.000  & 0.000  \\
$\Lambda_{\mathrm{s}}\mathrm{CDM}$       & 10.598 & -3.646 & -2.158 & -2.812 \\
L$\Lambda$CDM         & 10.455 & -1.788 &  1.820 & -0.122 \\
SSCDM                 & 10.392 & -1.852 &  1.757 & -0.185 \\
ECDM                  & 10.392 & -1.852 &  1.757 & -0.185 \\
\hline
\multicolumn{5}{c}{\textbf{CMB+BAO+SN}} \\
$\Lambda$CDM          & 1500.370 & 0.000 & 0.000  & 0.000  \\
$\Lambda_{\mathrm{s}}\mathrm{CDM}$  & 1489.347 & -9.023 & -9.014 & -3.574  \\
L$\Lambda$CDM         & 1487.836 & -8.535 & -8.514 & 2.363 \\
SSCDM                 & 1488.956 & -7.414 & -7.393 & 3.484 \\
ECDM                  & 1488.917 & -7.553 & -7.532 & 3.344 \\
\hline\hline
\end{tabular}
\caption{\justifying{\textit{Summary of the $\chi^2$ values and information criteria used for the comparison of cosmological models. The quantities $\Delta\mathrm{AIC}$, $\Delta\mathrm{AIC}_c$, and $\Delta\mathrm{BIC}$ are defined with respect to the $\Lambda$CDM model for each dataset combination and quantify the statistical preference of each model relative to $\Lambda$CDM.}}}
\label{tab:info_criteria_compact}
\end{table}

Turning to Table~\ref{tab:info_criteria_compact}, we compare the four models against $\Lambda$CDM using the AIC, AIC$_{\rm c}$, and BIC, as described in the previous subsection. According to Jeffreys’ scale (Table~\ref{deltaaic})\footnote{Combination I (CMB) is excluded from this analysis due to the insufficient number of data points for a reliable application of these criteria, which require $M \geq \kappa + 1$.}: 

\begin{itemize} 

\item Combination II (SN) moderately disfavours the sign-switching models, particularly under the BIC, where the stronger penalty for additional parameters results in all extensions being decisively disfavoured. This outcome reflects the limited constraining power of SN data on the additional parameters introduced by these models.

\item For the same reason, Combination III (BAO) also moderately disfavours all extensions, though more mildly, except for L$\Lambda$CDM and SSCDM, which are moderately disfavoured according to the AIC$_{\rm c}$ but deemed compatible with the BIC. Here, one can observe the difference between the values obtained from the original and corrected AIC. 

\item For Combination IV (CMB+BAO), a similar discrepancy is observed between the AIC and AIC$_{\rm c}$. Both criteria nevertheless support compatibility with $\Lambda$CDM for the smooth models, albeit with some penalisation associated with the additional parameter, while $\Lambda_{\mathrm{s}}\mathrm{CDM}$ remains moderately favoured. In contrast, the BIC finds all models to be compatible, with a slight preference for $\Lambda_{\mathrm{s}}\mathrm{CDM}$.

\item Combination V (CMB+BAO+SN) is favoured by the AIC, particularly in the case of the $\Lambda_{\mathrm{s}}\mathrm{CDM}$ model, owing to its relative simplicity. The AIC and AIC$_{\rm c}$ yield nearly identical results, reflecting the large statistical weight of the \textit{Pantheon+} and \textit{SH0ES} datasets. In contrast, the BIC moderately disfavors the smooth sign-switching models relative to the abrupt scenario, consistent with its stronger complexity penalty; i.e. its additional
parameter. Should tighter constraints be achieved, one would expect even stronger statistical support, particularly when employing the full \textit{Planck 2018} data combination, which we intend to explore in future work.

\end{itemize}

\begin{figure*}
    \centering
    \includegraphics[width=0.7\linewidth]{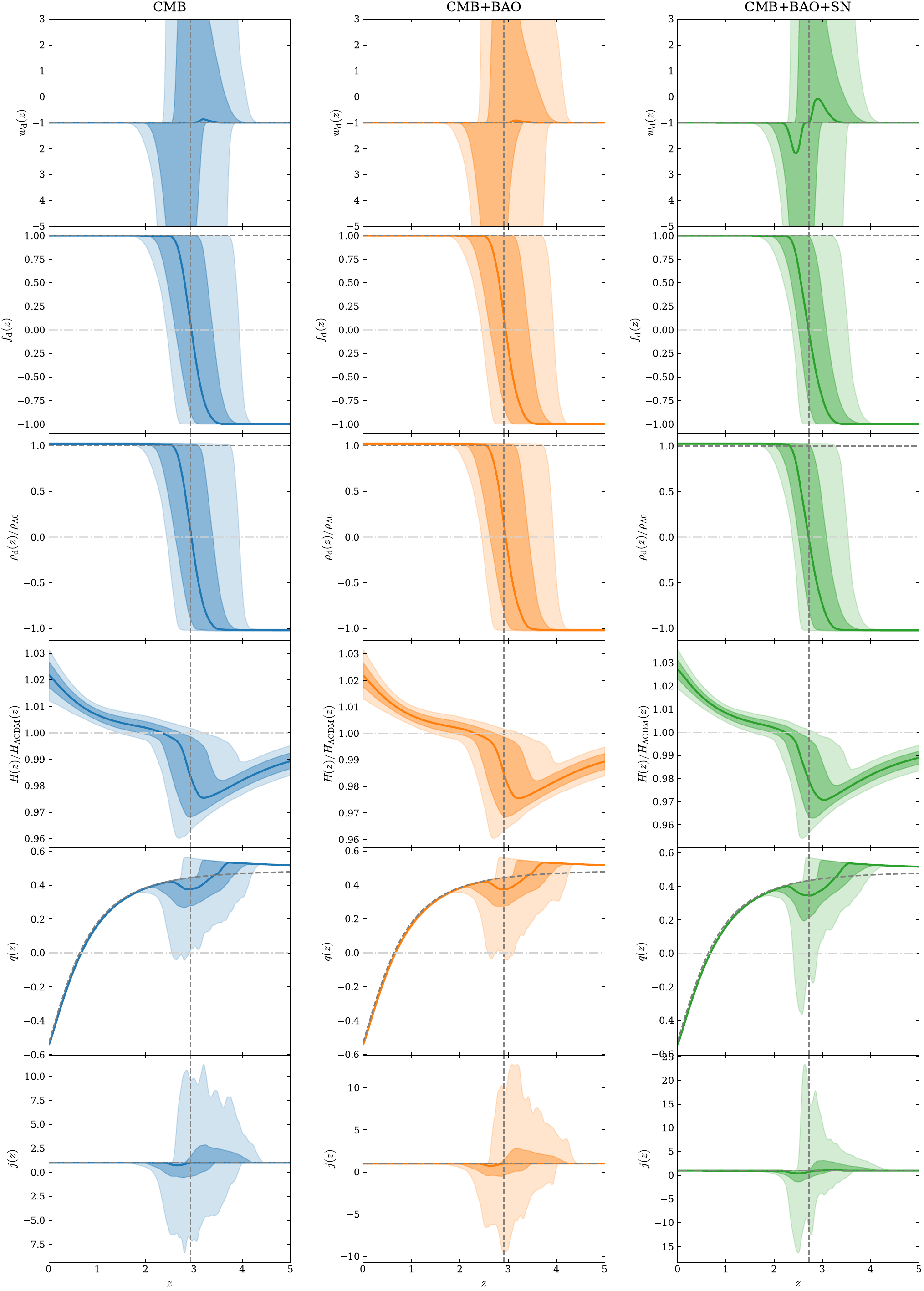}
    \caption{\justifying{\textit{Reconstructed EoS, energy density normalized to the current value, Hubble parameter, deceleration parameter and jerk for the SSCDM models with Planck18 + \textit{DESI DR2} + \textit{Pantheon+} $\&$ \textit{SH0ES} (left-most column), and with Planck18 + \textit{DESI DR2} (central column) and the Planck18 (right-most column). We also plot the reconstructed Hubble rate and energy density normalized to the Planck PR4 values for $\Lambda$CDM, for which $\Omega_{m0}$ = 0.315 and $H_0$ = 67.26 km/s/Mpc \cite{Rosenberg:2022sdy}. The solid colored lines represent the most-probable value and the shaded regions show the 68$\%$ and 95$\%$ confidence intervals around it. The grey dashed lines correspond to $\Lambda$CDM values. In the plots of $q(z)$ we also show in black dash-dotted line the border between deceleration ($q>$0) and acceleration ($q<$0) regimes, i.e., $q$=0.}}}
    \label{fig:sscdm-rec}
\end{figure*}

The results discussed above suggest that the lack of support for sign-switching models in Combinations II and III is primarily driven by the limited constraining power of these datasets on the additional parameters. As complementary datasets are combined, strengthening the overall constraints, sign-switching behaviour becomes progressively more supported, most notably in Combination V. While the abrupt $\Lambda_{\mathrm{s}}\mathrm{CDM}$ scenario is preferred over the more physically motivated smooth transitions, this preference largely reflects its relative simplicity. Achieving tighter constraints may therefore enhance the statistical viability of the smoother models.

It is note worthy that the SN data set, consisting of a large number of unbinned measurements directly probing the late-time expansion history, dominates the overall likelihood contribution. In contrast, the BAO data provide a small number of effective constraints, while the compressed \textit{Planck} likelihood encodes only a limited set of geometric parameters that are weakly sensitive to late-time modifications of the expansion history. As a result, the combined $\chi^2$ is necessarily driven to a large extent by the SN sector, which is expected in analyses that probe modifications of the late-time cosmological dynamics.

In this context, the improvement in $\chi^2$ between the $\Lambda$CDM and sign-switching models arises primarily from the supernova contribution and corresponds to a reduction of order $\mathcal{O}(10)$ over approximately $\mathcal{O}(10^3)$ data points, i.e. at the sub-percent level in terms of overall fit quality. While the AIC and BIC reflect this improvement and indicate a mild statistical preference for the sign-switching scenarios in some data combinations, we emphasise that these results should be interpreted with caution. In particular, the information criteria are largely driven by the data set with the highest effective constraining power (SN in this case), and should therefore be viewed as complementary rather than decisive indicators of model preference.

\subsection{Background and Cosmographic Reconstructions}

\begin{figure*}
    \centering
    \includegraphics[width=0.7\linewidth]{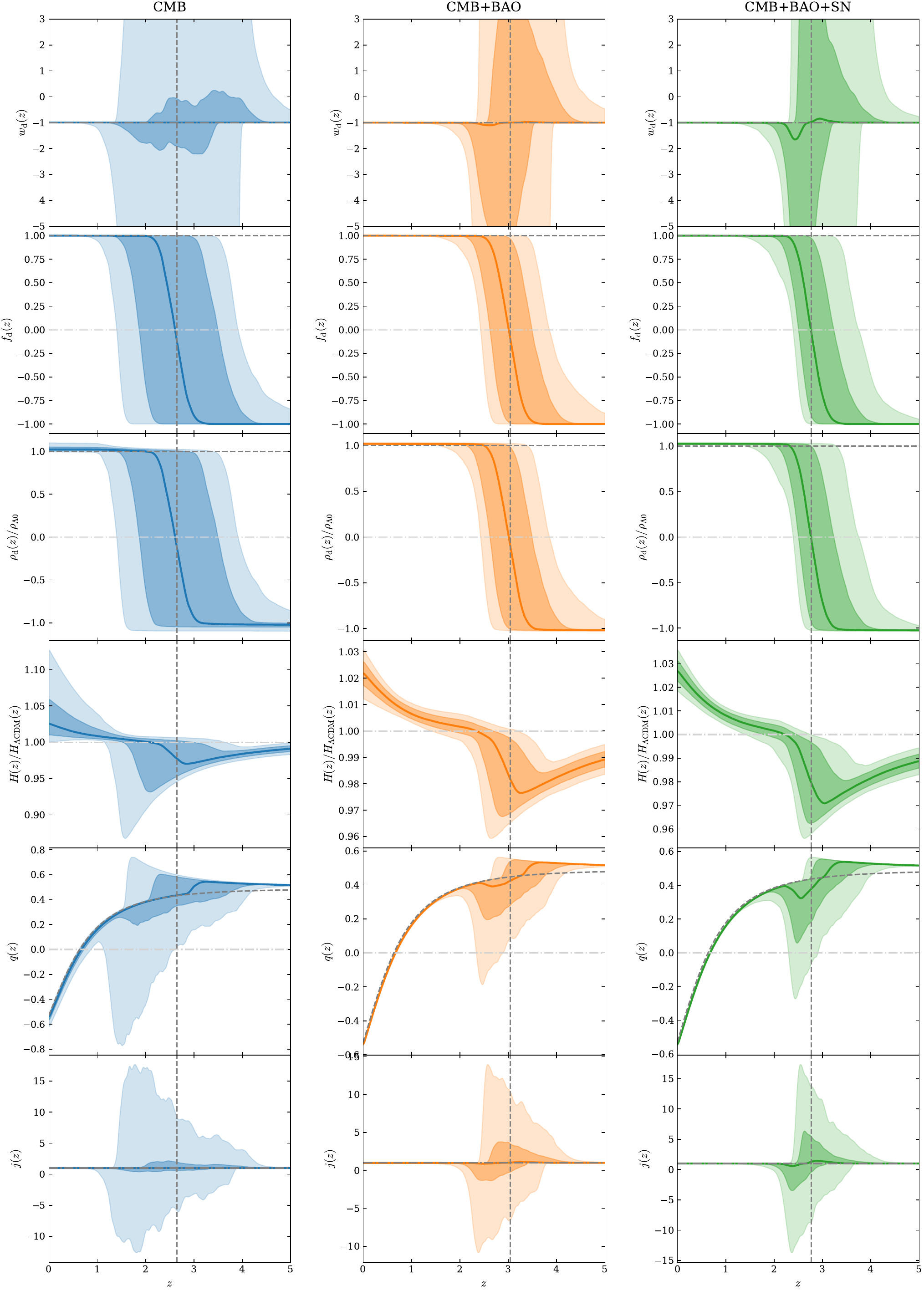}
    \caption{\justifying{\textit{Reconstructed EoS, energy density normalized to the current value, deceleration parameter and jerk for the ECDM model with Planck18 + \textit{DESI DR2} + PantheonPlus $\&$ SH0ES (left-most column), and with Planck18 + \textit{DESI DR2} (central column) and the Planck18 (right-most column). We also plot the reconstructed Hubble rate and energy density normalized to the Planck PR4 values for $\Lambda$CDM, for which $\Omega_{m0}$ = 0.315 and $H_0$ = 67.26 km/s/Mpc [114]. The solid colored lines represent the most-probable value and the shaded regions show the 68$\%$ and 95$\%$ confidence intervals around it. The grey dashed lines correspond to $\Lambda$CDM values. In the plots of $q(z)$ we also show in black dash-dotted line the border between deceleration ($q>$0) and acceleration ($q<$0) regimes, i.e., $q$=0.}}}
    \label{fig:ecdm-rec}
\end{figure*}

Lastly, in Figs.~\ref{fig:sscdm-rec} and~\ref{fig:ecdm-rec} we present the reconstruction of key background and cosmographic parameters for the SSCDM and ECDM models, respectively,\footnote{We perform the reconstruction only for the SSCDM and ECDM models, as these are the more physically motivated scenarios owing to their continuous evolution. This property has a significant impact on the behaviour of the EoS parameter, as well as on the deceleration and jerk parameters.
} from the present epoch up to redshift $z = 5$. These reconstructions build upon the analytical results derived in~\cite{Bouhmadi-Lopez:2025ggl}. In particular, we present the evolution of the DE equation-of-state parameter $w_{\mathrm{d}}(z)$, together with its energy density normalised to the present-day $\Lambda$CDM value, $\rho_{\mathrm{d}}(z)/\rho_{\Lambda0}$. We also show the corresponding dimensionless rescaled quantity $f_d(z)$ defined in Eq.~(\ref{def:f_d}), as well as the Hubble parameter $H(z)$ normalised to the $\Lambda$CDM value $H_{\Lambda CDM}(z)$ . In addition, we include the deceleration parameter $q(z)$ and the jerk parameter $j(z)$, thereby providing a comprehensive characterisation of the expansion history and kinematic behaviour of both models over the redshift range considered.

\subsubsection{The SSCDM model}

For the SSCDM model, the EoS parameter remains close to $-1$, exhibiting a significant deviation only during the sign-switching phase. From the evolution of $f_d(z)$ and $\rho_d(z)/\rho_{\Lambda0}$, we infer that the transition region is relatively broad for data Combinations I and IV, becoming significantly more constrained only in Combination V. From the evolution of $f_d(z)$ and $\rho_d(z)/\rho_{\Lambda0}$, we infer that the transition region is relatively broad for data Combinations I and IV, becoming significantly more constrained only in Combination V. We further observe that this model predicts a higher DE density than that inferred from \textit{Planck18} within the $\Lambda$CDM framework. An inspection of the ratio $H(z)/H_{\Lambda{\rm CDM}}(z)$ reveals that, prior to the sign-switching, the Hubble parameter is lower than the $\Lambda$CDM prediction, whereas after the transition it exceeds it in order to compensate, in agreement with the theoretical expectations of Refs.~\cite{Bouhmadi-Lopez:2025ggl,Bouhmadi-Lopez:2025spo}.

Regarding the deceleration parameter, one of the most intriguing implications of sign-switching models emerges, as anticipated in previous theoretical work: the possibility of a \textit{third phase of accelerated expansion} (see Fig.~3 of \cite{Bouhmadi-Lopez:2025ggl}). In the distant past, the deceleration parameter exceeds that of $\Lambda$CDM, since the effect of a negative DE density resembles that of an additional matter component. However, as the DE density increases, the deceleration parameter can cross the $q(z)=0$ threshold, thereby initiating a new phase of acceleration. It subsequently becomes positive again until the DE component dominates, ultimately leading back to the present accelerated expansion of the Universe. This behaviour was theoretically anticipated and is now supported observationally at the $95\%$ confidence level, where the $q(z)=0$ threshold is crossed twice.

Nevertheless, this feature may also be a consequence of the limited ability to constrain both the sign-switching redshift and the transition speed. In principle, there are two mechanisms by which a crossing of $q(z)=0$ may occur: either the transition is sufficiently fast, or the sign-switching redshift is low enough that, even with a slower transition, the threshold can still be crossed. Although our results suggest that such a possibility exists, more robust constraints are required before drawing firm conclusions. Whether a third phase of accelerated expansion is observable or not would provide a powerful test for accepting or ruling out this class of models and for constraining their dynamics. Finally, the jerk parameter also exhibits oscillatory features around the sign-switching epoch, which, if observable, could further shed light on the viability of these models.

\subsubsection{The ECDM model}

The ECDM model exhibits a qualitatively similar phenomenology to SSCDM, and we therefore refrain from repeating the full discussion. In this case, however, the transition redshift is better constrained, which leads to more clearly defined dynamical features. In particular, the crossing of the deceleration parameter threshold $q(z)=0$ appears more strongly favoured, reinforcing the possibility of an intermediate phase of accelerated expansion. Furthermore, the oscillatory behaviour of the jerk parameter around the sign-switching epoch is more pronounced in ECDM, suggesting that higher-order kinematical diagnostics may provide a clearer observational signature for this class of models.

\section{Conclusions}\label{sec4}

In this work, we have investigated a class of DE models featuring sign-switching behaviour, focusing on both abrupt and smooth transitions and confronting them with a variety of cosmological datasets, including CMB, BAO, and SN observations, as well as their combinations. Our analysis has shown that sign-switching models can provide a viable and, in some cases, preferred alternative to the standard $\Lambda$CDM framework when sufficient constraining power is available.

When individual datasets are considered, such as SN or BAO alone, the sign-switching models are generally disfavoured by information criteria. However, this result is primarily driven by the limited ability of these datasets to constrain the additional parameters introduced by the models, rather than by an intrinsic tension with the data. As the constraining power increases through combined analyses, particularly in Combination V (CMB+BAO+SN), sign-switching behaviour becomes increasingly favoured. In this context, the abrupt $\Lambda_{\mathrm{s}}\mathrm{CDM}$ model is often statistically preferred, mainly due to its relative simplicity and smaller number of free parameters. Nevertheless, the smooth sign-switching models exhibit comparable goodness-of-fit and display richer phenomenology, which may become statistically favoured as constraints improve.

From a dynamical perspective, both SSCDM and ECDM exhibit behaviour that departs non-trivially from $\Lambda$CDM. In particular, the evolution of the Hubble parameter, the deceleration parameter, and higher-order kinematical quantities such as the jerk parameter reveal distinctive signatures associated with the sign-switching epoch. Our results indicate that a crossing of the deceleration parameter threshold $q(z)=0$, corresponding to a potential intermediate phase of accelerated expansion, is possible within the allowed parameter space. This feature appears more clearly favoured in the ECDM model, where the transition redshift is better constrained, and where oscillations in the jerk parameter are more pronounced. If observable, such signatures could provide a powerful means of testing and potentially falsifying this class of models.

An important outcome of this study is the strong prior dependence exhibited by the sign-switching parameters, particularly the transition redshift $z_\dagger$ and the transition speed. We find that extending the prior range of $z_\dagger$ tends to push the preferred transition to higher redshifts, largely driven by the \textit{DESI DR2} three-dimensional BAO data, which do not allow for a negative DE density between their effective redshift points. Conversely, imposing tighter and more physically motivated priors brings the results closer to $\Lambda$CDM-like behaviour, suggesting that current constraints are not yet sufficient to robustly determine the detailed properties of the transition.
A similar conclusion applies to the transition speed parameter.

Overall, our results demonstrate that sign-switching DE models remain a promising and viable extension of $\Lambda$CDM, capable of alleviating cosmological tensions and producing distinctive dynamical signatures. At the same time, they highlight the crucial role of prior choices and dataset selection in shaping the inferred constraints. Future work will focus on revisiting the analysis with tighter and better-motivated priors, extending the explored parameter space for both $z_\dagger$ and the transition speed, and incorporating additional datasets, such as transverse BAO measurements and the full \textit{Planck 2018} likelihood. These improvements will be essential to robustly assess the physical viability of sign-switching DE and to determine whether its distinctive predictions can be tested observationally.

\section*{Acknowledgements}

The authors are grateful to \"{O}zg\"{u}r Akarsu, Hsu-Wen Chiang, Carlos G. Boiza, and  Nihan Kat{\i}rc{\i}  for discussions and insights on the current project. M. B.-L. is supported by the Basque Foundation of Science Ikerbasque. 
Our work is supported by the Spanish Grant PID2023-149016NB-I00 funded by (MCIN/AEI/10.13039/501100011033 and by “ERDF A way of making Europe). The authors acknowledge the contribution of the COST Action CA21136 “Addressing observational tensions in cosmology with systematics and fundamental physics (CosmoVerse)”. This work is also supported by the Basque government Grant No. IT1628-22 (Spain).

\bibliography{bibliography}

\end{document}